\documentclass[%
 reprint,
 amsmath,amssymb,
 aps, pre
]{revtex4-2}

\usepackage{graphicx}
\usepackage{dcolumn}
\usepackage{bm}
\allowdisplaybreaks
\usepackage{newtxtext}
\usepackage{cancel}
\usepackage[dvipsnames]{xcolor}
\usepackage[normalem]{ulem}
\usepackage{orcidlink}
\usepackage{placeins}
\hypersetup{
    colorlinks=true,
    citecolor=Cerulean,
    linkcolor=RubineRed,
    urlcolor=blue
}

\newcommand{\bbE}[1]{\mathbb{E}\left[#1\right]}
\newcommand{\bbV}[1]{\mathbb{V}\left[#1\right]}
\newcommand{\bbC}[2]{\mathbb{C}\left[#1,#2\right]}
\newcommand{\Var}[1]{\text{Var}[#1]}
\newcommand{\Ave}[1]{\langle#1\rangle}
\newcommand{\Cov}[1]{\text{Cov}[#1]}
\newcommand{\vsel}[0]{v_s}
\newcommand{\vmut}[0]{v_\mu}

\newcommand{\ser}[1]{\sum_{#1}}
\newcommand{\nser}[2]{\sum_{#1 \ne #2}}
\newcommand{\ntriser}[3]{\sum_{#1 \ne #2 \ne #3}}
\newcommand{\lrpar}[1]{\left(#1\right)}
\newcommand{\lrbra}[1]{\left[#1\right]}

\begin{document}

\preprint{APS/123-QED}

\title{Quasilocalization under coupled mutation-selection dynamics}

\author{C. J. Palpal-latoc\orcidlink{0000-0002-9171-6252}}
\email{cbpalpallatoc@up.edu.ph}
\affiliation{
 Core Facility for Bioinformatics, Philippine Genome Center, University of the Philippines System, Quezon City 1101, Philippines
}
\affiliation{
 National Institute of Physics, University of the Philippines Diliman, Quezon City 1101, Philippines
}

\author{Ian Vega\orcidlink{0000-0002-0428-048X}}
\affiliation{
 National Institute of Physics, University of the Philippines Diliman, Quezon City 1101, Philippines
}

\date{\today}

\begin{abstract}
When mutations are rampant, quasispecies theory or Eigen's model predicts that the fittest type in a population may not dominate. Beyond a critical mutation rate, the population may even be delocalized completely from the peak of the fitness landscape and the fittest is ironically lost. Extensive efforts have been made to understand this exceptional scenario. But in general, there is no simple prescription that predicts the eventual degree of localization for arbitrary fitness landscapes and mutation rates. Here, we derive a simple and general relation linking the quasispecies' Hill numbers, which are diversity metrics in ecology, and the ratio of an effective fitness variance to the mean mutation rate squared. This ratio, which we call the localization factor, emerges from mean approximations of decomposed surprisal or stochastic entropy change rates. On the side of application, the relation we obtained here defines a combination of Hill numbers that may complement other complexity or diversity measures for real viral quasispecies. Its advantage being that there is an underlying biological interpretation under Eigen's model. 
\end{abstract}

\maketitle

\section{Introduction}

Under canonical models in population genetics, mutation is seldom by design. The key contribution of quasispecies theory \cite{Eigen1971} (otherwise known as Eigen's model or the Eigen-Schuster model) is to reveal the full dynamical impact of mutation when it is very frequent, which is characteristic of asexual organisms such as bacteria but especially viruses \cite{Sniegowski2000, Sanjuan2016}. Whereas selection alone may lead to a ``survival of the fittest,'' quasispecies theory predicts that if mutation rates are high enough, less fit variants or ``types'' can even thrive because they are replenished through frequent mutation. The equilibrium population structure is called the quasispecies \cite{Nowak1992}. This provided a theoretical framework for understanding viral evolution and made a lasting influence in virology, where the term ``quasispecies'' is now used broadly to refer to the heterogeneous nature of viruses \cite{Domingo2002, Domingo2019}.

That less-fit mutants can persist did not originate from Eigen. Crow and Kimura have earlier studied a deterministic mutation-selection model \cite{CrowKimura1970}. But Eigen's model is defined for arbitrary number of types, frequencies and mutation rates. Under this generalization, the mass of the population can be arbitrarily distributed among the fittest and its mutant relatives. In extreme cases, the quasispecies may be dominated by a single type (localization) such as in survival of the fittest or dispersed over many or all types (delocalization). A well-studied example of the latter is the error catastrophe. Although there is no universal definition for it, it is often taken to be a (usually sharp) phase transition wherein the fittest type is lost from the population when its per-site mutation probability is above an ``error threshold'' \cite{Tarazona1992, Eigen2002}. Crossing this threshold, the population is totally delocalized from the peak of the fitness landscape (Fig. \ref{fig:qstheory}b). Vigorous theoretical effort has been spent understanding this threshold assuming mostly single-peaked fitness landscapes  \cite{Leuthausser1987, Galluccio1997, Saakian2006}, with extensions to drift \cite{Nowak1989, Alves1998, Campos1999}, recombination \cite{Boerlijst1996, Jacobi2006}, gene networks \cite{Tannenbaum2004a, Tannenbaum2004b} and epistasis \cite{HerreraMart2024}. Complementary to the error threshold, a total localization threshold has also been obtained that determines when the population is localized entirely on a fitness peak (i.e. survival of the fittest) \cite{McCaskill1984a}. The analogy between quasispecies theory and Anderson localization has also been discussed in Ref. \cite{Waclaw2011}. More recently, a noise-induced ``fidelity catastrophe'' has also been shown to occur wherein the population is always localized but switches dominant types over time \cite{Crosato2025}.

Despite these investigations, our understanding of localization for error-prone replication remains incomplete even in the deterministic limit. The particular value of the error or the localization threshold depends on the choice of simplified landscape and mutation scheme. In particular, it has been shown that the error threshold does not generically exist \cite{Wiehe1997, Baake2001, Bull2005, Wilke2005, Takeuchi2007}. Moreover, while error catastrophe and survival of the fittest carry important implications, they are extreme situations. Real error-prone populations should more commonly exist in an intermediate state of ``quasilocalization,'' where the population is diffused in the landscape around some dominant type, which might be the fittest. Indeed, this is the defining characteristic of real viral quasispecies \cite{Domingo2019, Domingo1978}, though we have poor analytical statements about this intermediate scenario. 

We understand that quasilocalization surely represents a balance between mutation and selection effects, with neither dominating. But can we say something more concrete and quantitative, especially beyond single-peaked landscapes? As a step forward, we use information-theoretic bounds to study the degree of localization analytically. Recently, speed limits on the time-evolution of observables, which may or may not be biological, have been obtained without relying on the particular details of the system \cite{Ito2018, Ito2020, Nicholson2020, Zhang2020, Adachi2022, Hoshino2023, GarcaPintos2024}. We take specifically the bounds in Ref. \cite{Hoshino2023} based on decomposing the surprisal or stochastic entropy change rate. Adapting them to Eigen's model, we derive an equilibrium relation for the degree of localization (Eq. \ref{eq:log_relation}). This relation links what we are calling the localization factor, which is computed from the statistics of the dynamical parameters, and the Hill numbers, which are canonical diversity metrics in ecology, of the quasispecies. 

This paper is organized as follows. In Sec. \ref{sec:model}, we introduce our formulation of quasispecies theory and important terminology of relevant quantities, including our definition of an ``influx rate.'' We also briefly review the speed limits in Ref. \cite{Hoshino2023} based on decomposing the surprisal rate. We apply them to Eigen's model, identifying localization and delocalization speeds.  In Sec. \ref{sec:results}, we obtain mean approximations of these speeds. We then derive our main results, the equilibrium relation involving the localization factor (Eq. \ref{eq:log_relation}) and its critical value. In Sec. \ref{sec:discussion}, we discuss the implications of our findings for the maintenance of information. We also propose a viral complexity index based on our results. Finally, we conclude in Sec. \ref{sec:conclusion}. 

The code for reproducing the figures and simulations in this work is available for download at \cite{palpallatoc}. This work benefited from open-source packages \texttt{matplotlib} \cite{matplotlib}, \texttt{numpy} \cite{numpy}, \texttt{scipy} \cite{scipy}, and \texttt{pandas} \cite{pandas2020, pandas2010}.

\section{\label{sec:model}Mutation-selection dynamics}

\subsection{Quasispecies theory}\label{sec:qstheory}

\begin{figure}
    \centering
    \includegraphics[width=\linewidth]{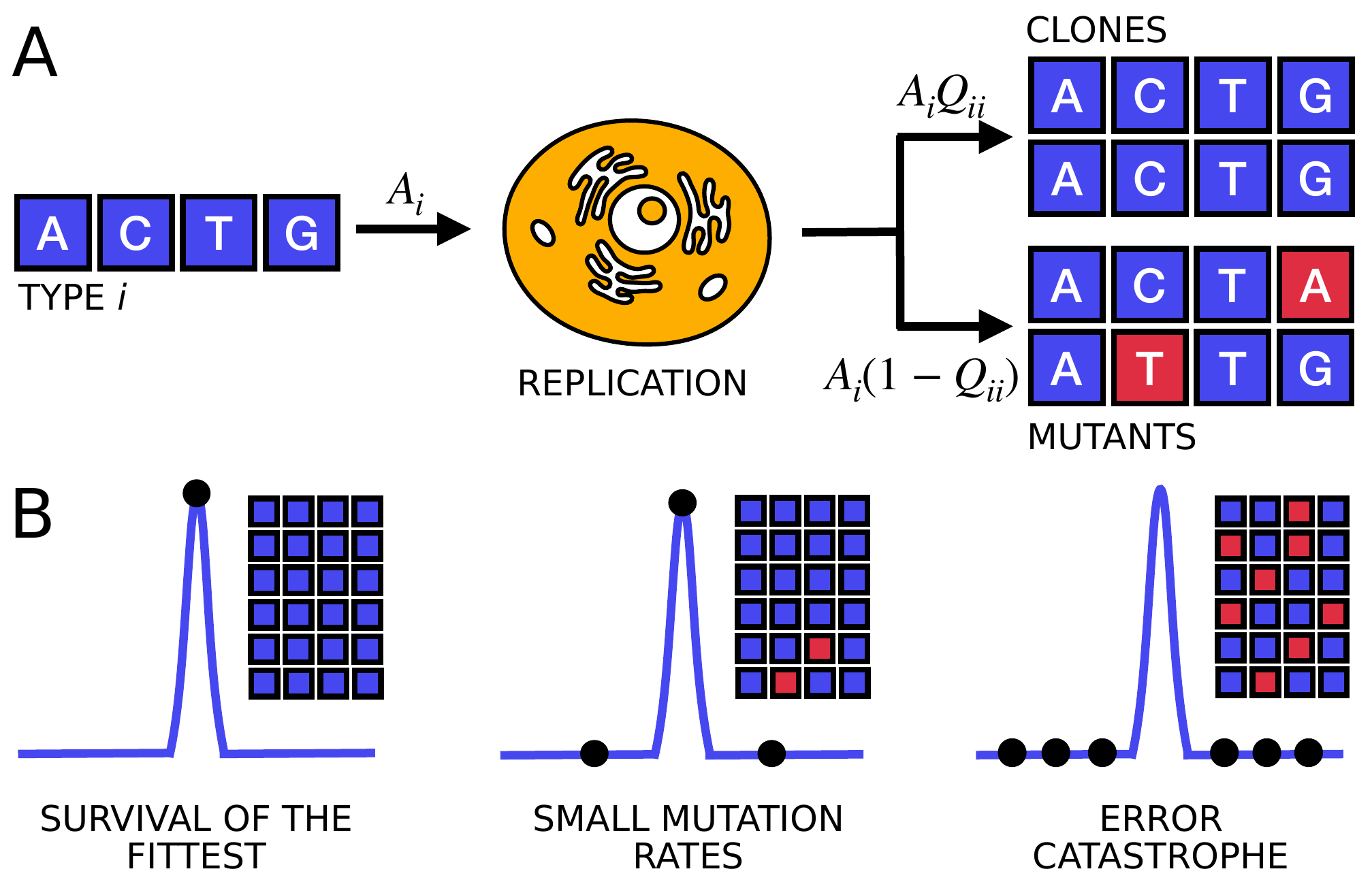}
    \caption{(A) In Eigen's model, mutants are produced as a result of erroneous replication, coupling the replication rate $A_i$ (hence, selection) of a type $i$, shown above as a sequence of nucleotides, and its copy fidelity $Q_{ii}$ (hence, mutation). (B) Much of the theoretical work has focused on the error catastrophe (right) in which the population (closed black circles) is delocalized from the peak of the fitness landscape (purple curve). On the opposite extreme is survival of the fittest (left) where the entire population is on the peak. The middle diagram represents the case when mutation is seldom.}
    \label{fig:qstheory}
\end{figure}

We begin by reviewing quasispecies theory, hereafter referred to as Eigen's model in which mutation and selection are coupled (Fig. \ref{fig:qstheory}a). When the total (census) population size $n_\text{tot}(t)$ is very large (technically, $n_\text{tot}(t) \rightarrow \infty$), a deterministic treatment applies \footnote{Real viral populations start out with a very small population size after undergoing a population bottleneck at the start of an infection. In this early stage, stochastic effects dominate and Eigen's model therefore does not apply. But unless the infection is cleared early on, the population size will eventually grow very large and the deterministic limit is virtually attained. For example, each person was estimated to carry $10^9-10^{11}$ copies of viral genetic material during peak infection of SARS-CoV-2 \cite{Sender2021}.}. For a population with $N \in \mathbb{N}$ different relevant types, the growth rate of a type $i \in \{1,\ldots, N\}$, is given by \cite{Eigen1971}
\begin{equation}\label{eq:qstheory}
    \dfrac{d}{dt}p_i(t) = (A_iQ_{ii} - D_i)p_i(t) + \sum_{j\ne i}^NA_jQ_{ij}p_j(t) - \Ave{E(t)} p_i(t),
\end{equation}
where $p_i(t)$ is the relative or normalized frequency of type $i$ and $\Ave{E(t)} := \sum_{i=1}^N\sum_{j=1}^NA_jQ_{ij}p_j(t)$ ensures the frequencies sum to one \footnote{In Eigen's original formulation, the total population size is \textit{constant} so that the frequency $p_i(t)$ in Eq. \ref{eq:qstheory} is \textit{absolute} and $\sum_{i=1}^N p_i(t) = n_\textit{tot}(t) = \text{constant}$. But Eq. \ref{eq:qstheory} also holds for a population with a \textit{time-varying} total population size with the frequency $p_i(t)$ being \textit{relative} and $\sum_{i=1}^N p_i(t) = 1$. Indeed, Eq. \ref{eq:qstheory} is exact for the mean relative frequencies of the unconstrained population (see Equation 2.12 of Ref. \cite{McCaskill1984b}) and the constant-size population is retrieved with the additional requirement that the covariance of the absolute frequencies vanishes (see Equation 3.14 of Ref. \cite{McCaskill1984b}). In either case, Eq. \ref{eq:qstheory} applies in the deterministic limit.}.  Here, the ``types'' we consider are implicitly haplotypes or sequences of nucleotides (Fig. \ref{fig:qstheory}a) comprising a portion or the entirety of the genetic material \footnote{Nothing however prevents applying Eq. \ref{eq:qstheory} to model evolving phenotypes or a more abstract population, biological or otherwise.}.

In Eq. \ref{eq:qstheory}, $A_i \in \mathbb{R}^+$ is the replication rate of type $i$. These replication rates define a one-dimensional ``fitness landscape,'' which is the mapping $f_A: i \mapsto A_i$ (Fig. \ref{fig:qstheory}b). (We adopt the ``landscape'' term for any other such mapping here.) When we speak of the fittest type or fitness peak, we are therefore referring to the type with the maximum replication rate. On the other hand, $Q_{ij} \in [0, 1]$, satisfying  $\sum_{j=1}^N Q_{ji} = 1$, is the probability of producing type $i$ during an erroneous replication of type $j$ when $j\ne i$ and the ``copy fidelity'' $Q_{ii}$ is the probability of accurate replication of type $i$. Because mutation is not negligible, it is not necessarily the type with the highest replication rate that will dominate a population.  Lastly, $D_i \in \mathbb{R}_{\ge 0}$ is a degradation variable. In the literature, $D_i$ is mostly ignored as it can be subsumed by $A_i$ but we keep this variable for completeness. 

Eq. \ref{eq:qstheory} is nonlinear due to $\Ave{E(t)}$, but it can be transformed into a linear problem that has an analytical solution \cite{Jones1976}. If we define the elements of what we will call the ``growth matrix'' $\mathsf{W} \in \mathbb{R}^{N\times N}$ as $W_{ij} := A_jQ_{ij} - D_i\delta_{ij}$,  where $\delta_{ij}$ is the Kronecker delta,  then the solution to Eq. \ref{eq:qstheory} is the weighted sum of the eigenvectors of $\mathsf{W}$, i.e. \cite{Thompson1974, Jones1976}
\begin{equation}\label{eq:qssolution}
    p_i(t) = \dfrac{\sum_{j=1}^Nc_jh_{ij}e^{\lambda_jt}}{\sum_{i=1}^N\sum_{j=1}^Nc_jh_{ij}e^{\lambda_jt}},
\end{equation}
where the coefficients $c_j$ are determined from initial conditions and $h_{ij}$ denotes the $i$th component of the $j$th eigenvector associated with $\lambda_j$. 

If $\mathsf{W}$ is nonnegative and irreducible, i.e. there is a mutational path from each type $i$ to any type $j$, then the Perron-Frobenius theorem \cite{Perron1907, Frobenius1912} guarantees that $\mathsf{W}$ has a positive, real, and maximum eigenvalue $\lambda_\text{max}$ and the components of its corresponding (right) eigenvector $\mathbf{h}_\text{max} := (h_{1\text{max}},\ldots, h_{N\text{max}})^T$ are all strictly positive too. The equilibrium steady-state, the ``quasispecies,'' is then determined purely from  $\lambda_\text{max}$ and $\mathbf{h}_\text{max}$:
\begin{equation}
    \lim_{t\rightarrow\infty} p_i(t) = \dfrac{h_{i\text{max}}}{\sum_{j=1}^Nh_{j\text{max}}} =: p_i^*.
\end{equation}
We refer to the existence of such a steady-state as the Perron-Frobenius property. This property may still be obeyed by some matrices that have negative entries (e.g. due to large $D_i$) as long as the off-diagonal entries are strictly nonnegative (i.e. $\mathsf{W}$ is a so-called Metzler matrix \cite{Arrow1989}) as is the case here. In this work, we assume that the Perron-Frobenius property is always satisfied. Then, when we speak of a type going ``extinct,'' we only mean that it attains a very small frequency.

\subsection{Influx rate}

\begin{figure}
    \centering
    \includegraphics[width=\linewidth]{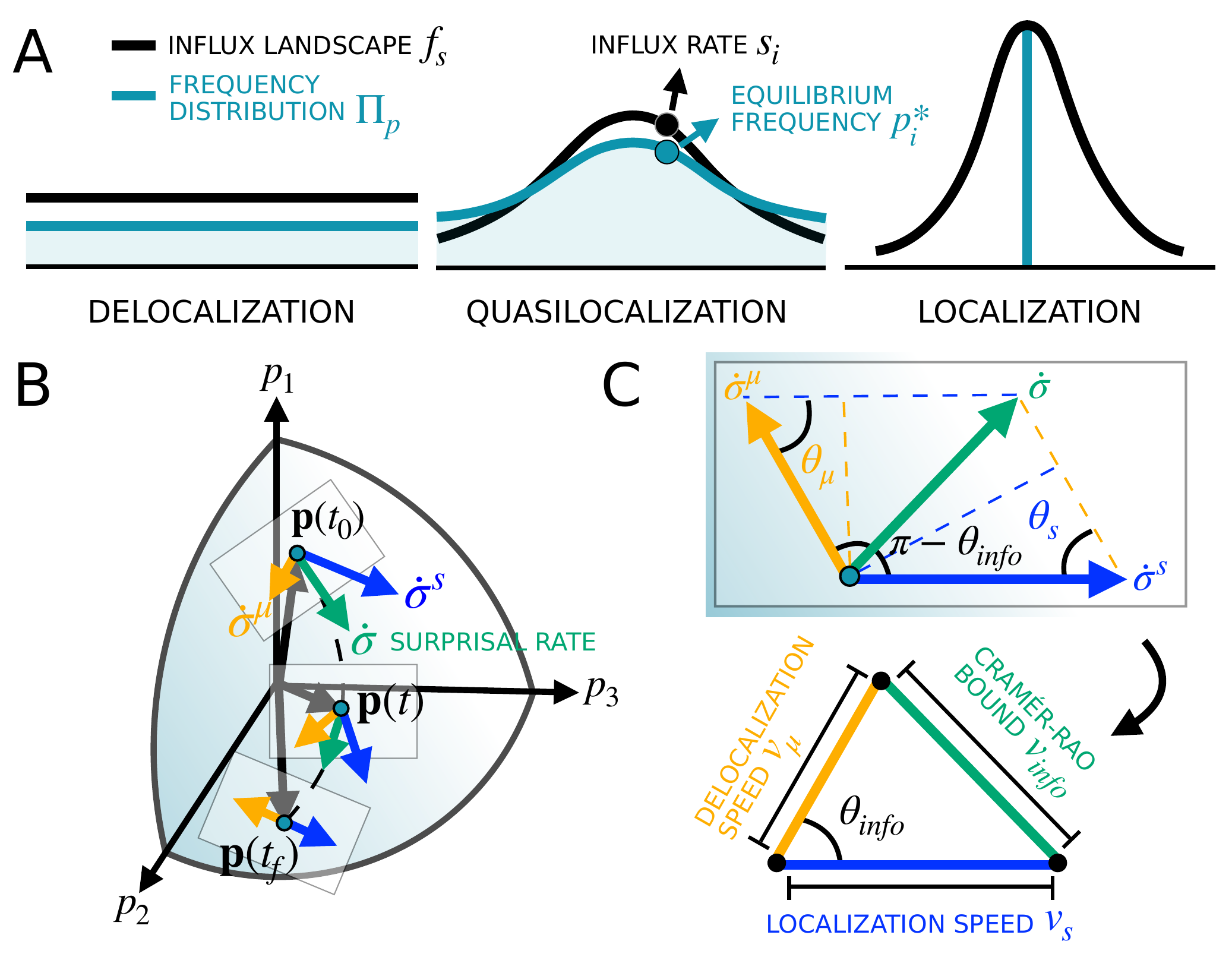}
    \caption{(A) Cartoon of the spectrum of equilibrium frequencies (cerulean curve) against the influx landscape (black curve). In the localized regime, the population concentrates on the peak of the influx landscape. In the delocalized regime, the population disperses over all types. (B) The population dynamics can be represented as motion along an octant of an $N$-dimensional hypersphere (or equivalently, as motion along an $N$-simplex). Here, $N=3$. Each point on this hypersphere corresponds to a vector of type frequencies $\mathbf{p}(t) \in \mathbb{R}^N$ that represents the state of the population. The population moves on the hypersphere from $t=t_0$ to $t=t_f$ with a net velocity given by the surprisal rate $\dot{\sigma}(t)$, which can be decomposed into contributions of the influx rate ($\dot{\sigma}^s$(t)) and mutation rate ($\dot{\sigma}^\mu$(t)). (C) The norm of these surprisal rates yield the (de)localization speeds and the Cramér-Rao bound.}
    \label{fig:hypersphere}
\end{figure}

Survival of the fittest and error catastrophe are almost synonymous with localization and delocalization, respectively. But the latter two can be defined without reference to fitness and in terms of the quasispecies distribution alone. Here, we take total localization to mean the dominance in equilibrium frequency of a single type, which may or may not be the fittest. That is, one type has frequency $p_i^* = 1$ and the rest have $p_{j\ne i}^* = 0$. On the other hand, we consider total delocalization as the complete absence of any dominance among types. That is, all types have the same frequency $p_i^* = 1/N$. The quasilocalized regime then corresponds to the intermediate case where types have differential frequencies. 

Based on these definitions, the diagonal entries of $\mathsf{W}$ should be the clear contributors to localization. More concretely, the eigenvalue equation for $\lambda_\text{max}$ is
\begin{equation}\label{eq:eigenvalue}
    \sum_{j=1}^N W_{ij}h_{j\text{max}} = \lambda_\text{max}h_{i\text{max}}.
\end{equation}
The dominance of a single type $i$ (i.e. $h_{j\text{max}} = 0 \ \forall j\ne i$) then implies from above that $W_{ii} = \lambda_\text{max}$. On the other hand, second-order perturbation theory gives \cite{Eigen1971}
\begin{equation}
    \lambda_i = W_{ii} + \sum_{j\ne i}^N \dfrac{W_{ij}W_{ji}}{W_{ii}-W_{jj}} + \ldots,
\end{equation}
which shows clearly that the off-diagonal entries  move the eigenvalues away from the diagonal. These entries intuitively should then contribute to delocalization. However, the delocalized limit of equal frequencies cannot be reached even if all $W_{ii}$ are the same. If the mutation rates are realistically heterogeneous, then types which benefit from more frequent mutation from other types will also have larger frequencies. In this situation, the delocalized limit can only be attained if the diagonals vary and compensate for the biases.

To systematically retrieve the localized and delocalized limits while allowing for mutational biases, we introduce the following transformation:
\begin{equation}\label{eq:influx_gauge}
    W_{ii} := s_i - \sum_{j\ne i}^NW_{ij},
\end{equation}
where $s_i \in \mathbb{R}$ is what we are calling the \textit{influx rate}. It is the sum of the rate of clonal replication of type $i$ (i.e. $W_{ii}$) and the mutational bias towards $i$ (i.e. $\sum_{j\ne i}^N W_{ij}$). In other words, $s_i$ is an effective ``total'' fitness parameter of type $i$. 

When these influx rates have the same value $s$, the dynamics becomes equivalent to
\begin{equation}
    \dfrac{d}{dt}n_i(t) = sn_i(t) -\sum_{j\ne i}^NW_{ij}n_i(t) + \sum_{j\ne i}^NW_{ij}n_j(t).
\end{equation}
To show that this has a delocalized solution, we need to solve the eigenvalue problem
\begin{equation}\label{eq:W_prime}
    \mathsf{W'}\mathbf{h}_\text{max} = (\lambda_\text{max}-s)\mathbf{h}_\text{max},
\end{equation}
where $\mathsf{W'}$ has the same off-diagonal entries as $\mathsf{W}$ but has diagonal entries $-\sum_{j\ne i}^{N} W_{ij}$. The dominant eigenvector $\mathbf{h}_\text{max}$ in Eq. \ref{eq:W_prime} can be retrieved by considering $\mathsf{W'}$ alone since $s$ merely shifts the eigenvalue. So consider that the constant vector $\mathbf{1} := (1,\ldots, 1)^T$ satisfies
\begin{equation}
    \mathsf{W'}\mathbf{1} = \mathbf{0},
\end{equation}
 That is, $\mathbf{1}$ is an eigenvector of $\mathsf{W'}$ corresponding to a zero eigenvalue. By noting that $\mathsf{W'}$ is irreducible and applying the Perron-Frobenius theorem indirectly \footnote{Define the matrix $\mathsf{M} := \mathsf{W'} + k\mathbb{I}$, where $\mathbb{I}$ is the identity matrix and $k$ is some large enough constant such that $\mathsf{M}$ is positive. Note also that $\mathsf{M}$ is irreducible because $\mathsf{W'}$ is. Now, $\mathbf{1}$ is an eigenvector of $\mathsf{W'}$ and also of $\mathsf{M}$. The Perron-Frobenius theorem guarantees that $\mathsf{M}$ has a unique (up to a scalar multiple) positive eigenvector which corresponds to its largest eigenvalue $\lambda_\text{max}^\mathsf{M}$. For $\mathsf{M}$, $\mathbf{1}$ is that eigenvector. To get back to $\mathsf{W'}$, see that $\mathsf{W'}\mathbf{h} = (\mathsf{M} - k\mathbb{I})\mathbf{h} = (\lambda^\mathsf{M} - k)\mathbf{h}$, where $\mathbf{h}$ is an eigenvector of either matrix and $\lambda^\mathsf{M}$ is an eigenvector of $\mathsf{M}$. Thus, maximizing $\lambda^\mathsf{M}$ retrieves the maximum eigenvalue $\lambda^\mathsf{W'}$ of $\mathsf{W'}$. Substituting $\mathbf{1}$ for $\mathbf{h}$, we find that  $\lambda^\mathsf{W'}_\text{max} = \lambda^\mathsf{M}_\text{max}-k = 0$.}, it can be shown that zero is the maximum eigenvalue of $\mathsf{W'}$ and therefore $\mathbf{1} = \mathbf{h}_\text{max}$ in Eq. \ref{eq:W_prime}. Similar influx rates then result to the delocalized solution $p_i^* = 1/N \ \forall i$. Meanwhile, these rates directly vary the diagonals $W_{ii}$, which remain important at localization. By tuning $s_i$, the two opposite localization limits can be reached. The population therefore evolves or (de)localizes on the influx landscape $f_s: i \mapsto s_i$ (Fig. \ref{fig:hypersphere}a). 

\subsection{Surprisal rate decomposition}

Even though there is no notion of heat in this abstract dynamical system, a trajectory-dependent or stochastic entropy can nevertheless be defined \cite{Crooks1999, Qian2002, Seifert2005}. Ignoring jumps to other states, the stochastic entropy of the system at time $t$ is $\sigma_i(t) := -\ln{p_i(t)}$. The corresponding entropy change rate is  $\dot{\sigma}_i(t) = -\dot{p_i}(t)/p_i(t)$, where the dot denotes the time derivative. In information theory, this stochastic entropy is better known as the Shannon information or surprisal. And the time derivative is the surprisal rate. 

In Ref. \cite{Hoshino2023}, the surprisal rate of a similar model to Eq. \ref{eq:qstheory} was decomposed into the contribution of selection and mutation. These component surprisal rates were used to obtain bounds on the rate of change of a general observable. What is remarkable about the bounds in Ref. \cite{Hoshino2023} is they are tighter than the Cramér-Rao limit, which is a universal speed limit on the time-evolution of an arbitrary observable \cite{Ito2020}. The selection and mutation bounds outperform the Cramér-Rao bound whenever selection or mutation is dominant, respectively. Thus, the bounds can be said to quantify the strength of selection and mutation. Our basic premise here is that a suitable decomposition of $\dot{\sigma}_i(t)$ can also be used to quantify the strength of localization and delocalization.

To this aim, we first rewrite the growth matrix as \begin{equation}
    \mathsf{W} := \mathsf{S} + \mathsf{U}.
\end{equation}
Here, $\mathsf{S} := \text{diag}(s_1,\ldots, s_N)$ is the influx matrix. And $\mathsf{U}$ is simply the mutation rate matrix, with off-diagonal elements $\mu_{ij} := W_{ij}$ and diagonal entries $\mu_{ii} := -\sum_{j\ne i}^NW_{ij}$. The surprisal rate $\dot{\sigma}_i(t)$ can then be decomposed into
\begin{equation}\label{eq:decomposition}
    \dot{\sigma}_i(t) = \dot{\sigma}_i^s(t) + \dot{\sigma}_i^\mu(t),
\end{equation}
where the decomposed rates are defined as
\begin{subequations}
\begin{align}
    \dot{\sigma}_i^s(t) &:= -s_i + \Ave{s(t)},\label{eq:surprisal_loc}\\
    \dot{\sigma}_i^\mu(t) &:= -\sum_{j=1}^N \mu_{ij}\dfrac{p_j(t)}{p_i(t)} + \Ave{\mu(t)}\label{eq:surprisal_deloc},
\end{align}
\end{subequations}
with $\Ave{s(t)} := \sum_{i=1}^Ns_ip_i(t)$ and $\Ave{\mu(t)} := \sum_{i=1}^N\sum_{j=1}^N\mu_{ij}p_j(t)$. These surprisal rates comprise the vectors $\dot{\sigma}^s(t) := (\dot{\sigma}_1^s(t), \ldots, \dot{\sigma}_N^s(t))$ and $\dot{\sigma}^\mu(t) := (\dot{\sigma}_1^\mu(t), \ldots, \dot{\sigma}_N^\mu(t))$ that control the motion of the population on an $N$-dimensional state space (Fig. \ref{fig:hypersphere}b). The net velocity being $\dot{\sigma}(t) := \dot{\sigma}^s(t) + \dot{\sigma}^\mu(t)$. Following Ref. \cite{Hoshino2023}, we can define the inner product $\langle X, Y\rangle := \sum_{i=1}^Np_i(t)X_iY_i$ and the norm $||X|| := \sqrt{\langle X, X\rangle}$. Applying this to $\dot{\sigma}(t)$ and to Eqs. \ref{eq:surprisal_loc} and \ref{eq:surprisal_deloc}, we introduce $v_\text{info} := ||\dot{\sigma}(t)||$, $v_s(t) := ||\dot{\sigma}^s(t)||$,  and $v_\mu(t) := ||\dot{\sigma}^\mu(t)||$. These norms are visualized in Fig. \ref{fig:hypersphere}c.

The norm of the net velocity $v_\text{info} := ||\dot{\sigma}(t)||$, where we adopt Ref. \cite{Hoshino2023}'s labeling, is an information geometric speed limit known as the Cramér-Rao bound. Meanwhile, we identify the norms $v_s(t)$ and $v_\mu(t)$, which have units of inverse time, as the localization and delocalization speeds, respectively. If mutation is neglected and $\dot{\sigma}_i(t) = \dot{\sigma}^s_i(t)$, then the population will localize towards the type with the largest influx rate (since $W_{ii} = s_i$).  So $v_s$ can be interpreted as the speed at which the population localizes. If the influx rates are equal instead and $\dot{\sigma}_i(t) = \dot{\sigma}^\mu_i(t)$, then multiple types will coexist due to their mutational support to each other. As we have shown above, each type will have exactly the same frequency. Hence, $v_\mu(t)$ can be interpreted as the delocalization speed. 

\section{\label{sec:results}Results}

\subsection{\label{sec:mean_approx}Mean approximation of speeds}

To look at the speeds more closely and achieve our main result, we seek their mean approximations. Rather than consider all individual influx and mutation rates, we take their moments instead. Accordingly, we consider the influx rates $s_i$ as having been sampled from some distribution $\Pi_s$ and the mutation rates $\mu_{ij}$ as having been sampled from some distribution $\Pi_\mu$. Provided that $N$ is large, the expectation of $v_s$ and $v_\mu$ with respect to these distributions will approach their true values. This follows from the law of large numbers for the $N$-term sums in $v_s$ and $v_\mu$. The resulting quantities can be interpreted as the ensemble average of speeds of populations with statistically similar growth matrices.

In this approximation scheme, we assume that we can replace the rates with their moments. When calculating the expectations, we essentially take the frequencies in $v_s$ and $v_\mu$ to be independent from these rates. In the localized regime, and particularly at equilibrium, this does not hold because the types' frequencies will strongly covary with their influx and mutation rates. However, we are least interested in this regime since the population's equilibrium frequency distribution is already understood: the type with largest $s_i$ will dominate. Meanwhile, in the quasilocalized regime, the covariance between the rates and frequencies is weaker because the frequency is not solely determined by any particular rate. In the delocalized limit, the covariance is exactly zero because all frequencies are equal. 

Since our main regime of interest is the quasilocalized scenario, where the population's frequency distribution is most nontrivial, our approximation remains relevant. In addition, one of our results that the localization factor $F$ (Sec. \ref{sec:eq_relation}) controls the degree of localization carries into the localized regime. It is only how this factor relates to the equilibrium frequencies exactly that our approximation becomes inapplicable. We shall discuss more about the approximation error and its implication for our main result later.

Lastly, we also assume, outside the delocalized limit, that no two types are dynamically equivalent. That is, no two types share the same influx rate and mutational support, though some rates may be arbitrarily close but not exactly the same \footnote{This is not unreasonable despite synonymous mutations that produce the same protein (and hence, supposedly the same fitness effect). In the case of SARS-CoV-2, for example, there is still variation in the fitness effects of synonymous mutations \cite{Bloom2023} and the site-specific mutation rates are even more variable \cite{Haddox2025}.}. When there is variation among the influx and mutation rates, the rates vary smoothly. That is, the variance of the rates is not driven by outliers or by separated clumps. This assumption still includes a wide class of realistic influx and mutation landscapes.

Taking the mean of the speeds (squared), we obtain (SM Secs. \ref{app:loc_speed} and \ref{app:deloc_speed})
\begin{subequations}\label{eq:approx_speeds}
\begin{align}
    \bbE{(v_s)^2} &= \bbV{s}\left(1-\sum_{i=1}^Np_i^2\right),\label{eq:approx_loc_speed}\\
    \begin{split}
        \bbE{(v_\mu)^2} &= \bbE{\mu}^2\lrpar{\sum_{i=1}^N\dfrac{1}{p_i}-N^2}\\
        &\quad+ \bbV{\mu}\lrbra{\sum_{i=1}^Np_i^2\lrpar{\sum_{j=1}^N\dfrac{1}{p_j}-2N}-N+2},
\end{split}\label{eq:approx_deloc_speed}
\end{align}
\end{subequations}
where we denote by $\bbE{x}$ and $\bbV{x}$ the mean and variance of $x$ with respect to its distribution $\Pi_x$ as opposed to the distribution of equilibrium frequencies $\Pi_p$. We also suppress the time dependence of quantities from now on. 

\subsection{Behavior of the speeds}

Let us discuss Eq. \ref{eq:approx_loc_speed} first. It is intuitive that the influx rate variance emerges as the only relevant parameter in the localization speed. The dominance of one type will depend on how large (and therefore, different) its rate is compared to the other types'. The sum of the squared frequencies is also instructive. This sum has a maximum possible value of one, which it attains only when one type dominates. That is, the speed becomes zero when the population is already localized. Using the QM-AM inequality, the sum has a minimum value of $1/N$, which it attains if $p_i=1/N$. The speed is therefore maximal if the population is delocalized. When it is already localized, the speed is zero.

This behavior of localization can be seen by returning to the ``microscopic'' equations. Denote by $p_\textit{m}$ the frequency of the type with the largest influx rate $s_\text{max}$ and let the remaining types retain the usual notation. Then the dynamics, in the mutation-free case, is governed by
\begin{equation}
    \dfrac{d}{dt}\ln p_m = s_\text{max} - \Ave{s}, \ \dfrac{d}{dt}\ln p_i =s_i -\Ave{s},
\end{equation} 
with $s_\text{max} > s_i \ \forall i$. When a type reaches dominance,  $\Ave{s} \approx s_\text{max}$. The growth rate of the dominant type vanishes, and the rest go to extinction. That is, localization is complete. 

Moving on to Eq. \ref{eq:approx_deloc_speed}, we notice that the delocalization speed is more sensitive to small frequencies due to the $1/p_i$ terms. A single type on the brink of extinction can drive the speed to blow up. This makes sense since under delocalization, there is no dominating type which would require other types to go extinct. If a type is found at a small frequency, perhaps as an initial condition, the population must move quickly to avoid its extinction under delocalization. On the other hand, defining the convex function $g(x):= 1/x$, Jensen's inequality with equal weights of one yields $N^2 \le \sum_{i=1}^N1/p_i$. That is, the smallest value of $\sum_{i=1}^N1/p_i$ occurs if all $p_i=1/N$. This drives both the mean and variance terms to zero. The variance terms in particular in Eq. \ref{eq:approx_deloc_speed} are generally smaller because of the $p_i^2$ factors of $\sum_{j=1}^N 1/p_j$ (cf. $\mathcal{O}(1)$ of the mean terms) and $N$ term (cf. $N^2$). For the rates we consider in this work, we have also checked numerically that they do have negligible impact. 

The critically low frequency $p_i = 1/N$ can  be seen from the microscopic dynamics. Suppose that all types have the same influx rate, then the dynamics is governed purely by mutation rates:
\begin{equation}\label{eq:mutation_only}
    \dfrac{dp_i}{dt} = \sum_{j=1}^N\mu_{ij}p_j - \Ave{\mu}p_i.
\end{equation}
From above, it is clear that a type's frequency modulates its growth rate, which decreases if it is larger and increases if it is smaller. If we replace $\mu_{ij}$ by the mean mutation rate $\mathbb{E}[\mu]$, then we have $- \Ave{\mu}p_i \approx 0$. Eq. \ref{eq:mutation_only} then becomes
\begin{align}
    \dfrac{dp_i}{dt} &= \mu_{ii}p_i + \sum_{j\ne i}^N\mu_{ij}p_j,\\
    &= -\sum_{j\ne i}^N\mu_{ij}p_i + \sum_{j\ne i}^N\mu_{ij}p_j,\\
    &= \mathbb{E}[\mu](1-p_i) + \mathbb{E}[\mu](-N+1)p_i,\\
    &= \mathbb{E}[\mu](1 -Np_i).
\end{align}
Thus, whether a type will grow or shrink depends on whether its current frequency is above or below $1/N$.

\subsection{\label{sec:eq_relation}Equilibrium relation}

Having a firmer grasp of the speeds, we now focus on our main result. Our aim is to to derive a relationship between the parameters of the system and a function of frequencies that quantifies localization. We do not seek a particular form of the latter as there is no unique way to measure localization. Our only requirement is that this function must monotonically vary as the population becomes more (de)localized. With this clarified, our starting point is that the speeds should contain the answer. It is intuitive that if $v_s$ ($v_\mu$) is much greater than $v_\mu$ ($v_s$) throughout the dynamics, then the population localizes (delocalizes). Hence, the relative dominance between these speeds should determine the degree of localization. We shall return to this point later (Sec. \ref{sec:critical_factor}). Instead, we notice that these speeds equal each other at equilibrium. Since the surprisal rate $\dot{\sigma}$ of the population must vanish then, it follows that $\dot{\sigma}^s$ and $\dot{\sigma}^\mu$ should have the same norm to cancel each other (Fig. \ref{fig:hypersphere}b). This can be shown explicitly by taking the mutation speed and substituting 
\begin{equation}
    \sum_{j\ne i}^N\mu_{ij}h_{j\text{max}} = (\lambda_\text{max} - W_{ii})h_{i\text{max}}
\end{equation}
from Eq. \ref{eq:eigenvalue} to remove the mutation rates. After expanding the squares, the result is (SM Sec. \ref{app:equal_speeds})
\begin{equation}\label{eq:equal_speeds}
    v_\mu^2 = \sum_{i=1}^Np_i^* s_i^2 - \left(\sum_{i=1}^Np_i^*s_i\right)^2 = v_s^2.
\end{equation}

Because of Eq. \ref{eq:equal_speeds}, we can write the following equality using the approximate speeds:
\begin{equation}\label{eq:equal_approx_speeds}
    \bbE{v_s^2} + \mathcal{C}_s(\mathbf{p}^*) = \bbE{v_\mu^2} + \mathcal{C}_\mu(\mathbf{p}^*),
\end{equation}
where $\mathbf{p}^* := (p_1^*,\ldots, p_N^*)$ is the vector of equilibrium frequencies, and  $\mathcal{C}_s(\mathbf{p}^*)$ and $\mathcal{C}_\mu(\mathbf{p}^*)$ are the covariance corrections to $\bbE{v_s^2}$ and $\bbE{v_\mu^2}$, respectively, that become significant in the localized regime. We have derived $\mathcal{C}_s(\mathbf{p}^*)$ explicitly as (SM Sec. \ref{app:correction_terms})
\begin{equation}
    \begin{split}
        \mathcal{C}_s(\mathbf{p}^*) &:= N\lrpar{\bbC{s^2}{p^*} -\bbC{s^2}{(p^*)^2}} + N\bbC{s}{p^*}^2\\ 
        &\quad-2\bbE{s}N\lrpar{\bbC{s}{p^*} - \bbC{s}{(p^*)^2}}\\
        &\quad-N^2\bbC{s}{p^*}^2,
    \end{split}
\end{equation}
which shows that the correction is indeed driven by covariance between the rates and frequencies.

Substituting the approximate speeds, we can rewrite Eq. \ref{eq:equal_approx_speeds} as
\begin{align}
    \dfrac{\bbV{s}}{\bbE{\mu}^2}   &= \mathcal{H}(\textbf{p}^*) + \dfrac{\mathcal{C}_\mu(\mathbf{p}^*)-\mathcal{C}_s(\mathbf{p}^*)}{\bbE{\mu}^2\lrpar{1-\sum_{i=1}^N(p_i^*)^2}},\label{eq:full_relation}\\
    \mathcal{H}(\textbf{p}^*) &:= \dfrac{\sum_{i=1}^N1/p_i^*-N^2}{1-\sum_{i=1}^N(p_i^*)^2}.\label{eq:hill_numbers}
\end{align}
where we dropped the negligible mutation variance terms in Eq. \ref{eq:approx_deloc_speed}. In SM Sec. \ref{app:correction_terms}, we show numerically that both covariance corrections, $\mathcal{C}_\mu(\mathbf{p}^*)$ and $\mathcal{C}_s(\mathbf{p}^*)$, are negative overall and become significant when the population enters the localized regime. 

We observe the following about Eq. \ref{eq:full_relation}. Its right-hand side diverges if the population is completely localized. The denominators of both terms approach zero. At the same time, the numerator of $\mathcal{H}(\textbf{p}^*)$ diverges and the covariance corrections on the second term become important. When the population is completely delocalized instead, both terms vanish as both numerators go to zero. In general, $\mathcal{H}(\textbf{p}^*)$ monotonically increases with increasing localization \footnote{For a convex function $g$, the sum $\phi := \sum_{i=1}^N g(x_i)$ is Schur-convex \cite{Marshall2010}. That is, $\phi$ increases as $\mathbf{x} := (x_1,\ldots, x_N)$ becomes more localized. Then, the sums $\sum_{i=1}^N 1/p_i^*$ and $\sum_{i=1}^N (p_i^*)^2$, which have convex addends, both increase with increasing localization, and as a result, so does $\mathcal{H}(\textbf{p}^*)$.}. The right-hand side of Eq. \ref{eq:full_relation} therefore quantifies the population's degree of localization and by equality, so does the left-hand side. More than this though, notice that the left-hand side contains only the statistics of the growth matrix.  We recognize that the covariance corrections in the right-hand side of Eq. \ref{eq:full_relation} also contains the influx and mutation rates. But the term involving these corrections still depends on the solution. Despite containing the rates, it is overall a response variable. Then, Eq. \ref{eq:full_relation} can be understood in the following way. The ratio 
\begin{equation}
    F := \bbV{s}/\bbE{\mu}^2,
\end{equation}
which we now call the \textit{localization factor}, is not just equal to but sets the degree of localization as measured by the right-hand side of Eq. \ref{eq:full_relation}, which can be approximated by $\mathcal{H}(\textbf{p}^*)$ outside the localized regime.

Eq. \ref{eq:full_relation} naturally teases the possibility of using either the frequencies or rate moments alone to infer or predict the other. But because of the covariance corrections, this can only be done very accurately deep in the quasilocalized regime or near the delocalized limit. Nevertheless, the order of $F$ should track well the order of $\mathcal{H}(\textbf{p}^*)$, especially outside the localized regime. In lieu of Eq. \ref{eq:full_relation}, we can write instead the following approximation where the corrections are suppressed:
\begin{equation}
    \log_b{F} \approx \log_b{\mathcal{H}(\textbf{p}^*)},\label{eq:log_relation}
\end{equation} 
in a chosen base $b$. Eq. \ref{eq:log_relation} fulfills our main goal of directly linking the parameters to the equilibrium frequencies. It shows that the degree of localization is the result of the strength of mutation against fluctuations in the \textit{influx} landscape, which mutation rates also contribute to. Eq. \ref{eq:log_relation} suggests that to fully localize a population, $\bbV{s}\rightarrow\infty$. To fully delocalize, $\bbV{s}\rightarrow 0$. The localized limit can also be reached when $\bbE{\mu}\rightarrow 0$ but only if the influx rates are still variable. The delocalized limit can also be reached when $\bbE{\mu}$ becomes extremely large provided that the influx rates are the same. Thus, at the opposite limits, $\bbV{s}$ already contains sufficient information and the mutation rates matter only in so far as they affect $\bbV{s}$. Between these limits, the mean mutation rate $\bbE{\mu}$ plays a more pronounced role curbing the localizing effect of $\bbV{s}$; a larger $\bbV{s}$ is required to localize when $\bbE{\mu}$ is increased. These implications are not at all surprising. Instead, the contribution of Eq. \ref{eq:log_relation} is to provide an approximate analytical relationship for general fitness landscapes and mutation rates.

\subsection{Numerical test}

We now test Eq. \ref{eq:log_relation}, which in all calculations here is evaluated in base $10$, by numerically obtaining the equilibrium frequencies for a spectrum of localization factors. For this purpose, we introduce three independent measures for the degree of localization: first, the Shannon entropy $S := -\sum_{i=1}^Np_i\ln{p_i}$, which goes to zero in the localized limit and $\ln{N}$ in the delocalized limit; second, the participation ratio $PR := \sum_{i=1}^N p_i^2$ which is a classic measure of localization and goes to one in the localized limit and $N$ in the delocalized limit; and lastly, we also introduce $R_{50}$ which is similar to $PR$. We define $R_{50}$ to be the minimum number of types needed to cover at least $50\%$ of the population. This metric satisfies $1 \le R_{50} \le N/2$. Just like $PR$, $R_{50}$ gives an effective number of types that contribute the most frequency to the population. We mention $R_{50}$ in addition to $PR$ because the former defines a natural threshold for the localization regime, $R_{50}=1$. For each localization factor, it also varies less than $PR$. Note that these three metrics are inversely proportional to the degree of localization, attaining their highest value in the delocalized limit.

\begin{figure}
    \centering
    \includegraphics[width=\linewidth]{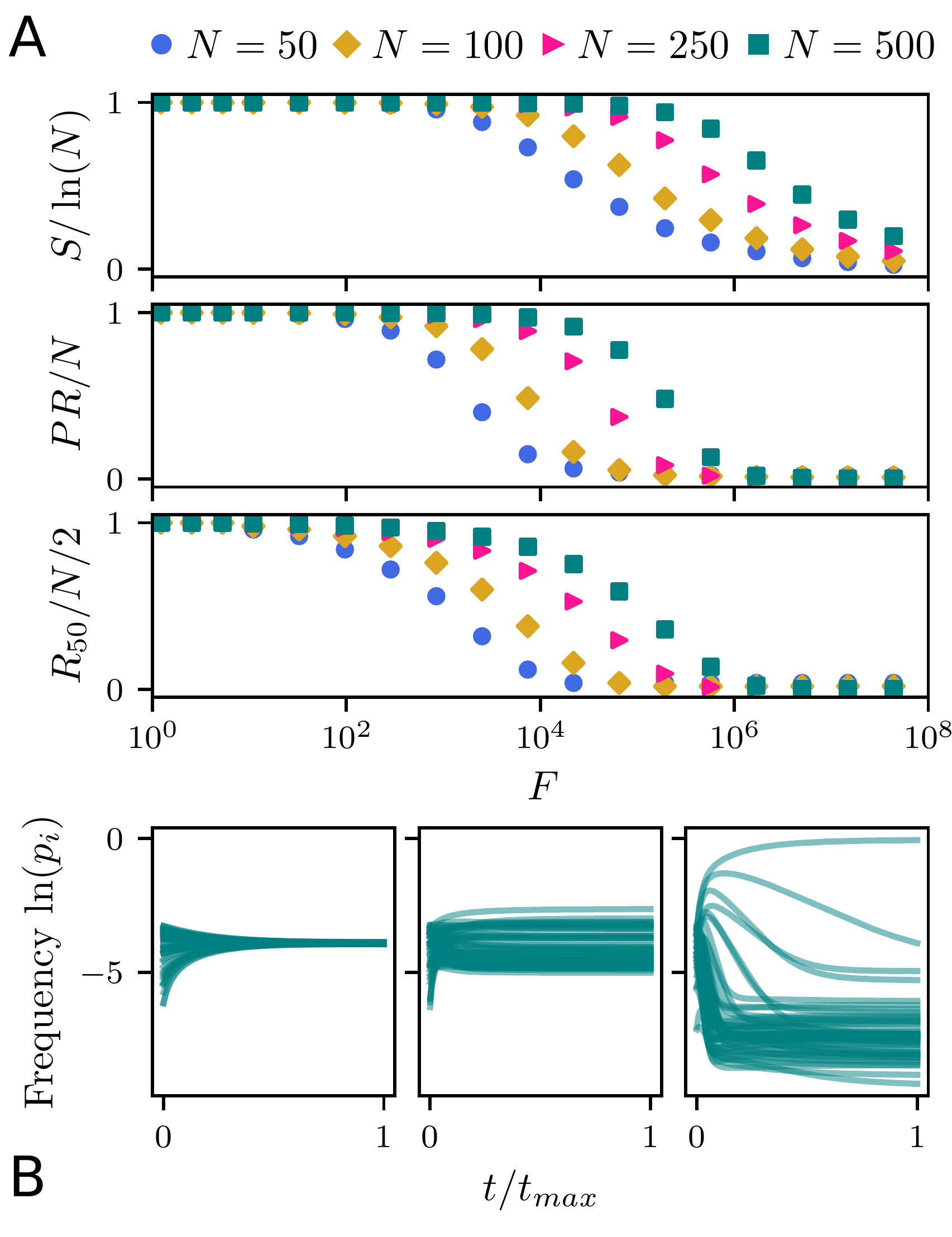}
    \caption{(A) The degree of localization as measured by different metrics increases with $F$. Each point is the median value of the respective metric over $10^3$ realizations per given $F$. Notably, a  population with larger $N$ will localize at a larger $F$. (B) Different $F$ lead to qualitatively different frequency dynamics. The left panel shows the time-evolution of frequencies at $F=1$, the middle at $F=1.5\times10^3$, and the right at $F=5.5\times10^5$, with $N=50$ in all cases. The frequencies are plotted against time normalized by the maximum time of integration $t_{max}$ (see Eq. \ref{eq:tmax} in main text).}
    \label{fig:degrees}
\end{figure}

In Fig. \ref{fig:degrees}a, we show that indeed the localization factor $F$ sets how localized the population is. The degree of localization, as measured through the median $S$, $PR$ and $R_{50}$, increases with increasing $F$. (Note that $R_{50}$ plateaus before reaching the delocalized and localized limits by definition.) Notably, we find that larger factors are needed to localize populations with larger $N$. This makes sense because larger $N$ means more mutational paths and more types that need to be suppressed in the localized regime. Fig. \ref{fig:degrees}b demonstrates directly that the magnitude of $F$ lead to qualitatively different dynamics, which agree with our expectations. Small $F$ (left panel of Fig. \ref{fig:degrees}b) leads to delocalization and large $F$ (right panel) leads to localization. An intermediate value (middle panel) quasilocalizes the population. In Fig. \ref{fig:degrees}b and all time-evolution plots in this work, the dynamics as given by Eq. \ref{eq:decomposition} is integrated until $t_\text{max}$ when equilibrium is effectively reached. This is defined as
\begin{equation}\label{eq:tmax}
    t_\text{max} := \ln{100}/(\lambda_\text{max} - \lambda_2),
\end{equation}
where $\lambda_2$ is the second-largest eigenvalue of $\mathsf{W}$. That is, $t_\text{max}$ is roughly the time it takes for the dominant mode, $e^{\lambda_\text{max}t}$, in Eq. \ref{eq:qssolution} to be $100$ times larger than the second-largest mode. 

\begin{figure}
    \centering
    \includegraphics[width=\linewidth]{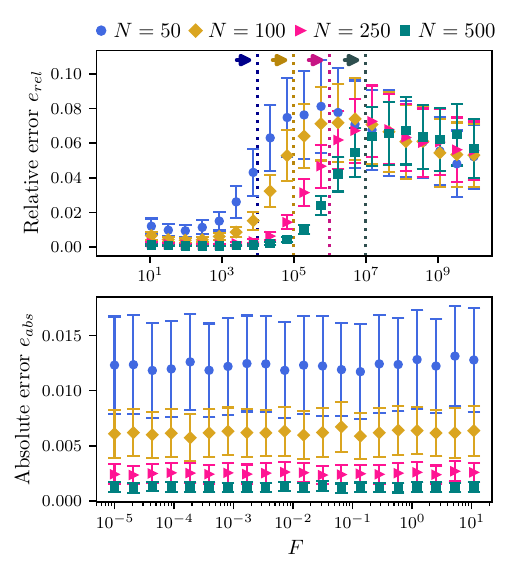}
    \caption{Accuracy of the equilibrium relation. See the main text for exact definition of relative and absolute errors. Top: The relation Eq. \ref{eq:log_relation} has a median relative error $e_\text{rel}$ within $10\%$ across regimes, with median errors below $6\%$ in the quasi- and delocalized regimes. The dotted lines and arrows (top of the plot) indicate the approximate onsets of the localized regimes for different $N$ that have the same but darker color scheme as the errors. Here, the onset of the localized regime is defined as when the median $R_{50}$ for a given $F$ drops to one. Bottom: The median absolute error $e_\text{abs}$ is shown for low values of $F$ as the relative error becomes undefined when $F \rightarrow 0$. Each median error in both panels is computed from $10^3$ realizations per given $F$. The $95\%$ confidence intervals are constructed from the interquartile range ($IQR$) of the computed errors: $\pm 1.57\times IQR/\sqrt{10^3}$. }
    \label{fig:errors}
\end{figure}

To obtain the accuracy of Eq. \ref{eq:log_relation}, we sample columns of $\mathsf{U}$--excluding the diagonals--from the Dirichlet distribution, each with random parameters, and then scale it by a random number drawn uniformly. This choice was made to mimic the constraint $\sum_{j=1}^N Q_{ji}A_i = A_i$. We then draw influx rates from a uniform distribution such that they have the variance to achieve the desired $F$. Fig. \ref{fig:errors}a shows the relative error $e_{rel}$, which we define as
\begin{equation}
    e_{\text{rel}}^2 :=  \text{median}\{(\log_{10}{\mathcal{H}_i(\textbf{p}^*)}/\log_{10}{F_i} -1)^2\},
\end{equation}
and Fig. \ref{fig:errors}b shows the absolute error, which we define as
\begin{equation}
    e_{\text{abs}}^2 :=  \text{median}\{(\log_{10}{\mathcal{H}_i(\textbf{p}^*)}-\log_{10}{F_i} -1)^2\}.
\end{equation}

Looking at Fig. \ref{fig:errors}b first, we see clearly that the absolute error improves as $N$ increases which is expected since our approximation is only valid for large $N$. In Fig. \ref{fig:errors}a, we find that within the quasilocalized regime, the median relative errors are generally below $6\%$ and get smaller as the population becomes more delocalized (i.e. smaller $F$). In the localized regime (to the right of the dotted lines in Fig. \ref{fig:errors}a), the median error notably rises, falls and appears to plateau around $6\%$. In SM Sec. \ref{app:num_corrections}, we show numerically that this occurs because the approximation error or covariance correction $\mathcal{C}_s(\mathbf{p}^*)$ rises initially faster than $\mathcal{C}_\mu(\mathbf{p}^*)$. These corrections, which have the same sign, eventually attain similar orders and cancel each other incompletely, causing the error to fall later at large $F$. We also tested Eq. \ref{eq:log_relation} using other mutation rate distributions and find similar results (SM Sec. \ref{app:numerical_implementation}).

\subsection{\label{sec:critical_factor}Critical localization factor}

\begin{figure}
    \centering
    \includegraphics[width=\linewidth]{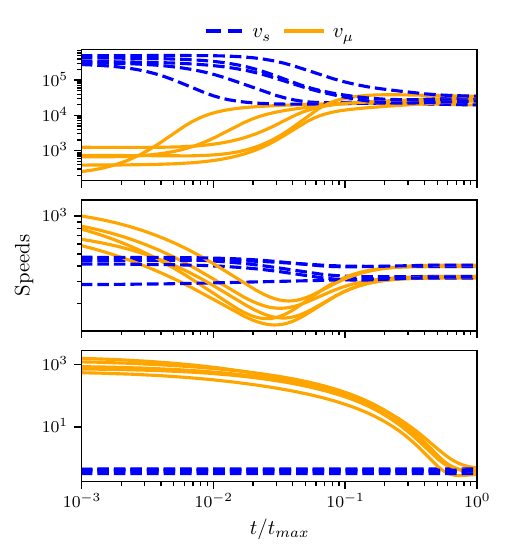}
    \caption{Time-evolution of pairs of localization and delocalization speeds for $F = 1.5\times 10^6$ (top panel), $F = 1.5\times 10^3$ (middle), and $F = 1.5$ (bottom) with $N = 50$ in all cases. If $v_s$ ($v_\mu$) starts out much larger, it stays larger than $v_\mu$ ($v_s$) until equilibrium. If the speeds have similar orders, they may switch in dominance (middle panel) over time but they will still have comparable magnitudes.}
    \label{fig:speeds}
\end{figure}

The localization factor is akin to a control parameter which can be varied to access different localization phases of the population. We can ask whether there is a critical value of $F$ such that the population undergoes a localization phase transition after crossing it. To this end, we point out that the relative dominance between the speeds is maintained until equilibrium when they equal each other. Since the dynamics is deterministic, the population must always be converging toward the steady-state within the time limit given by $1/\lambda_\text{max}$. That is to say, if the quasispecies is (de)localized, the population must have been (de)localizing throughout until equilibrium. We show numerical examples in Fig. \ref{fig:speeds}. If $v_s$ starts out much larger than $v_\mu$ or vice versa, this relative dominance between them is preserved in the nonequilibrium phase. If the speeds however start out approximately equal, they may switch in dominance but neither gets significantly larger than the other.

Because this relative dominance is maintained, we can set one speed to be greater at the start of the dynamics and determine $F$ such that the population will (de)localize from its initial state. Given a vector of initial frequencies $\mathbf{p}^\dagger := (p_1^\dagger, \ldots, p_N^\dagger)$ and using Eqs. \ref{eq:approx_loc_speed} and \ref{eq:approx_deloc_speed}, we obtain the following conditions and their resulting effect to the population:
\begin{subequations}\label{eq:phase}
\begin{align}
F \gg \mathcal{H}(\textbf{p}^\dagger) &\rightarrow \text{localize},\label{eq:phase_loc}\\
F = \mathcal{H}(\textbf{p}^\dagger) &\rightarrow \text{retain degree of localization},\label{eq:phase_retain}\\
F \ll \mathcal{H}(\textbf{p}^\dagger)  &\rightarrow \text{delocalize},\label{eq:phase_deloc}
\end{align}
\end{subequations}
where we did not need to be concerned with the covariance corrections since we are far from equilibrium. If $\mathcal{H}(\textbf{p}^\dagger)\rightarrow\infty$ or $\mathcal{H}(\textbf{p}^\dagger)\rightarrow 0$, then no value of $F$ can localize or delocalize the population even further, respectively. In which case, any finite $F$ moves the population away from either extreme limit. What is more interesting is the case of an initially quasilocalized. Then, $\mathcal{H}(\textbf{p}^\dagger)$ serves as a critical value of $F$. With a factor above or below $\mathcal{H}(\textbf{p}^\dagger)$, the population transitions to the localized or delocalized regime, respectively. The phase space of an initially quasilocalized population can be categorized into three regions, which are characterized by whether $F$ is greater than, less than, or equal to  $\mathcal{H}(\textbf{p}^\dagger)$. 

\begin{figure}
    \centering
    \includegraphics[width=\linewidth]{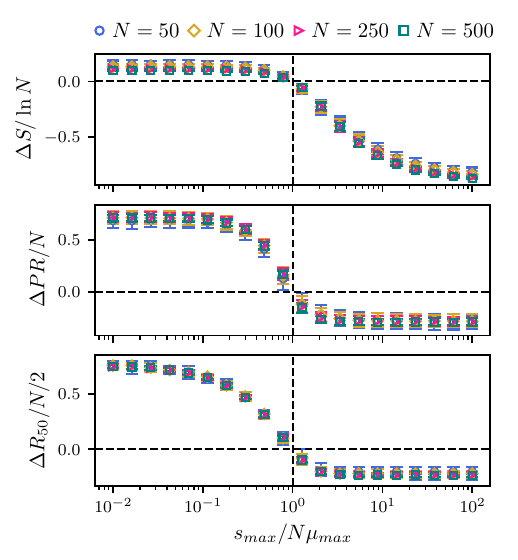}
    \caption{For uniformly distributed parameters and initial frequencies (see main text), $s_\text{max}/\mu_\text{max} = N$ marks a critical point from which the population (de)localizes. The above panels show the difference in degree of localization $\Delta X := X(\textbf{p}^*) - X(\textbf{p}_0^*)$ where $X$ is the localization metric ($S$, $PR$, or $R_{50}$), $\textbf{p}^*$ is the quasispecies evolved under $s_\text{max}/N\mu_\text{max} \ne 1$, and $\textbf{p}_0^*$ is the quasispecies evolved under $s_\text{max}/\mu_\text{max} = N$. Each point is the median value of $10^3$ realizations per $s_\text{max}/N\mu_\text{max}$. The narrow error bars shown mark the interquartile range ($IQR$). The $95\%$ confidence intervals for the median constructed from the $IQR$, $\pm 1.57\times IQR/\sqrt{10^3}$, are much tighter and not shown.}
    \label{fig:change}
\end{figure}

To gain more intuition, we can rewrite $\mathcal{H}(\textbf{p}^\dagger)$ for an initially quasilocalized population. First, we note that the denominator of $\mathcal{H}(\textbf{p}^\dagger)$ may be approximated by unity in this regime, especially for large $N$. Next, the numerator can be expressed in terms of the statistics of the absolute frequencies. Eq. \ref{eq:phase_retain} then becomes (SM Sec. \ref{app:uniform_dist_freq})
\begin{equation}\label{eq:phase_reciprocal}
    F \approx (N^2-N)\bbE{n^\dagger}\bbE{1/n^\dagger},
\end{equation}
where the dagger indicates the absolute frequencies are initial or far from equilibrium. Next, the  first-order Taylor series expansion of $\bbE{x/y}$ is given by \cite{Kendall2010}
\begin{equation}
    \bbE{x/y} \approx \dfrac{\bbE{x}}{\bbE{y}} - \dfrac{\bbC{x}{y}}{\bbE{x}^2} + \dfrac{\bbV{x}}{\bbE{y}^3}.
\end{equation}
Applying this to Eq. \ref{eq:phase_reciprocal}, we have
\begin{equation}\label{eq:phase_cv}
    F \approx (N^2-N)(CV_{n^\dagger}^2 + 1) \approx N^2CV_{n^\dagger}^2,
\end{equation}
where $CV_{n^\dagger}^2 := \bbV{n^\dagger}/\bbE{n^\dagger}^2$ is the squared coefficient of variation. Thus, it becomes harder to localize the population when there are many types, which we have already seen previously, but also when the frequencies have high variability.

As a concrete example, we mention a related and interesting numerical result  in Ref. \cite{Hoshino2023} (particularly, their Appendix C). The authors observed that their selection and mutation bounds equal in tightness whenever the maximum clonal replication rate ($\lambda_\text{max}$ in their notation) is $N$ times as large as the maximum mutation rate (their $\mathsf{W}_\text{max}$). There are subtle differences between their variables and ours. But we can get a similar result, following some of the assumptions made in Ref. \cite{Hoshino2023}. First, we can quantify the strength of localization and delocalization through
\begin{subequations}
\begin{align}
    \theta^s &:= \arccos\left(\dfrac{v_\text{info}-v_\mu}{v_s}\right),\\
    \theta^\mu &:= \arccos\left(\dfrac{v_\text{info}-v_s}{v_\mu}\right),
\end{align}
\end{subequations}
respectively (see Fig. \ref{fig:hypersphere}c), analogous to how Ref. \cite{Hoshino2023} quantified the strength of selection and mutation through their $\theta^*$ and $\theta^\dagger$ (Eqs. 15 and B3 of Ref. \cite{Hoshino2023}, respectively) \footnote{To be clear, they quantified the tendency of selection and mutation bounds to give tighter limits through their $\mathcal{P}^*(\theta^*)$ and $\mathcal{P}^\dagger(\theta^\dagger)$. But considering  $\theta^*$ and $\theta^\dagger$ directly is equivalent.}. Equal strength of (de)localization (analogous to equal tightness of the bounds in Ref. \cite{Hoshino2023}) is retrieved when $\theta^s = \theta^\mu$ ($\theta^* = \theta^\dagger$ in Ref. \cite{Hoshino2023}). It can be shown \footnote{Let $a = v_\text{info}$, $b = v_s$, and $c = v_\mu$. Then, $\theta^s = \theta^\mu \implies (a-b)/c = (a-c)/b \implies (b-c)(a - (b+c)) = 0$. This has two solutions: either $a = b + c$ or $b = c$. The former is a special case of the latter. After expanding $v_\text{info}$ and noting the identity $\sqrt{x} + \sqrt{y} = \sqrt{x + y + 2\sqrt{xy}}$, it can be written as $\Cov{\dot{\sigma}^s, \dot{\sigma}^\mu} = \sqrt{\Var{\dot{\sigma}^s}\Var{\dot{\sigma}^\mu}}$, where the covariance and variances are evaluated with respect to the type frequencies. This states that the norm of the projection of $\dot{\sigma}^\mu$ ($\dot{\sigma}^s$) onto $\dot{\sigma}^s$ ($\dot{\sigma}^\mu$) is equal to the norm of $\dot{\sigma}^\mu$ ($\dot{\sigma}^s$), which would imply that $\Var{\dot{\sigma}^s} = \Var{\dot{\sigma}^\mu}$ or $v_s = v_\mu$.} that this is equivalent to $v_s = v_\mu$, which is Eq. \ref{eq:phase_retain}.

\begin{figure}
    \centering
    \includegraphics[width=\linewidth]{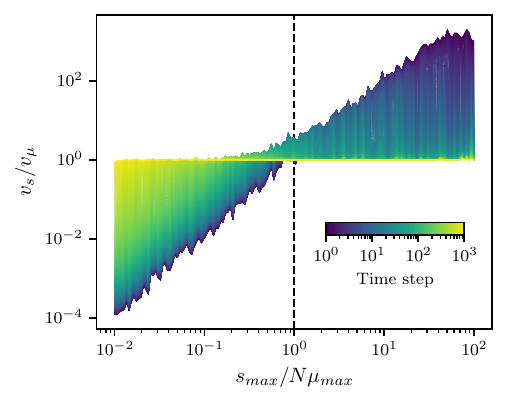}
    \caption{Time-evolution of $v_s/v_\mu$ over $10^3$ time points per given $s_\text{max}/N\mu_\text{max}$. Since the time of integration to reach equilibrium at $t_\text{max}$ (see Eq. \ref{eq:tmax} in main text) varies for different parameter values, each solid trace of the same color indicates the same time step (out of $10^3$ time points until $t_\text{max}$) instead rather than the same clock time of integration. Thus, the solid violet line which traces a ragged diagonal from $v_s/v_\mu \sim 10^{-4}$ to $v_s/v_\mu \sim 10^4$ is comprised by initial values of $v_s/v_\mu$ across $150$ different instances of $s_\text{max}/N\mu_\text{max}$. The solid yellow horizontal line, where $v_s/v_\mu = 1$, constitutes equilibrium at time $t_\text{max}$, which would be different for each $s_\text{max}/N\mu_\text{max}$. For uniformly distributed rates and initial frequencies, $v_s$ dominates until equilibrium (yellow line) whenever $s_\text{max}/N\mu_\text{max} \gg 1$ and $v_\mu$ dominates until equilibrium whenever $s_\text{max}/N\mu_\text{max} \ll 1$.}
    \label{fig:ratio}
\end{figure}

Next, we also let the absolute (initial or nonequilibrium) frequencies be drawn randomly from a uniform distribution with support ranging from $10^{x_1}$ to $10^{x_2}$, where $x_1,x_2 \in \mathbb{R}^+$ and $x_2 > x_1$. We also take the rates to be drawn from uniform distributions. Let $s \sim U(-s_\text{max}, s_\text{max})$ and $\mu \sim U(0, \mu_\text{max})$. Applying these assumptions to Eq. \ref{eq:phase_retain}, we get (SM Sec. \ref{app:uniform_dist_freq})
\begin{equation}
    \dfrac{s_\text{max}^2}{\mu_\text{max}^2} = (N^2-N)\dfrac{3}{4}\lrbra{\dfrac{\ln 10}{2}\dfrac{1 + 10^{-\Delta x}}{1 - 10^{-\Delta x}}\Delta x - 1},
\end{equation}
where $\Delta x := x_2-x_1$. 
The support of $n$ in Ref. \cite{Hoshino2023} was not explicitly mentioned but if the maximum difference between any frequency is two orders of magnitude ($\Delta x = 2$), we obtain
\begin{equation}\label{eq:max_N}
    \dfrac{s_\text{max}}{\mu_\text{max}} \approx N.
\end{equation}
In Fig. \ref{fig:change}, we verify that Eq. \ref{eq:max_N} acts like a critical point. For $s_\text{max}/N\mu_\text{max} > 1$, we find that the population localizes. On the other hand, when $s_\text{max}/N\mu_\text{max} < 1$, the population delocalizes. We have checked that $s_\text{max}/N\mu_\text{max} \approx 1$ yields a quasilocalized population. In Fig. \ref{fig:ratio}, we see that this can attributed to how $v_s$ and $v_\mu$ evolve over time. When $s_\text{max}/N\mu_\text{max}$ is much greater (less) than one, the speed $v_s$ starts out much larger (smaller) than $v_\mu$ and maintains its relative dominance over time until equilibrium. Note that in the quasilocalized scenario ($s_\text{max}/N\mu_\text{max} \approx 1$), there is still a type which has the largest frequency so that $v_s$ must still be somewhat larger than $v_\mu$ to quasilocalize the population.

\section{\label{sec:discussion}Discussion}

\subsection{\label{sec:information}Maintenance of information}

Information has been defined as a relative quantity representing correlation between two systems \cite{Adami2004}, e.g. a population and its environment. To be concrete, we take the following definition of information that a population stores about an environment as a difference in entropies \cite{Adami2002, Adami2016}:
\begin{equation}\label{eq:info_entropy}
    I_{\Delta S} := S_\text{max} - S(\mathbf{p}^*)
\end{equation}
where $S_\text{max}$ is the maximum uncertainty about the population and $S(\mathbf{p}^*) := -\sum_{i=1}^Np_i^*\log_b{p_i^*}$ is the entropy given that the population is in the environment. The maximum uncertainty is $S_\text{max} = \log_b{N} = L$ if the logarithm base is the alphabet size of the type sequence. Thus, the maximum information that we can maintain scales with $L$, which occurs when $S(\mathbf{p}^*)=0$ or when we know the environment prefers a single type.  Eigen's original error threshold is a statement about the maximum possible length $L_\text{max}$ (and hence, the maximum information content) that can be faithfully reproduced.

From Eq. \ref{eq:info_entropy}, it is clear that the degree of localization is directly tied to the possible amount of information that can be maintained. Our equilibrium relation Eq. \ref{eq:log_relation} and the conditions in Eq. \ref{eq:phase} align with previous findings \cite{Eigen1971, Watkins2008, Peck2010, Hledk2022} that mutation impedes preservation of information.  Indeed, one can view the critical value of $F$, Eq. \ref{eq:phase_retain}, as a generalized delocalization threshold for an initially quasilocalized population. It is akin to a ``soft'' error threshold. Eq. \ref{eq:phase_reciprocal} does not imply an error catastrophe where the fittest becomes extinct, which we recall is landscape-dependent. But Eq. $\ref{eq:phase_reciprocal}$ does prescribe when the structure of the quasispecies, and therefore the information shared with the environment through Eq. \ref{eq:info_entropy}, becomes lost.

We also mention a subtle point about how mutation can muddle information. Note that fitness (i.e. the replication rate $A_i$) is central to the concept of information as a correlation between a population and its environment. This is because that correlation is physically manifested through natural selection. (Mutation does not correlate with the environment provided that mutation rates are not directly selected for or environmental effects do not modulate mutation rates.) However, the population localizes on the influx landscape so that frequencies are not proportional to fitness in general. With Eq. \ref{eq:info_entropy}, we could obtain $I_{\Delta S} = 0$ even if there is variation in fitness when such variation compensates for mutational biases leading to delocalization. On the other hand, even if there is no variation in fitness, mutational biases can still lead to a quasilocalized population. In which case, we will have computed $I_{\Delta S} \ne 0$ even though selection did not take place. Only when the mutation probabilities $Q_{ij}$ keep the influx‑rate ranking of types identical to their fitness ranking can the information preserved, in the sense of Eq. \ref{eq:info_entropy}, faithfully reflect selection strength. We note that mutation is not unique in this aspect; genetic drift  can fix certain types even though they have no fitness advantage. 

Alternative to Eq. \ref{eq:info_entropy}, we can take inspiration from Ref. \cite{Hledk2022} and define information as the Kullback-Leibler divergence between $\mathbf{p}^*$ and its equivalent constant-fitness quasispecies:
\begin{equation}\label{eq:info_kl}
    I_{KL} := -S(\mathbf{p}^*) - \sum_{i=0}^Np_i^*\log_b{p_i^\mu},
\end{equation}
where $p_i^\mu$ denote the equilibrium frequencies of types sharing a replication rate (but with the same mutation probabilities that produced $\mathbf{p}^*$). This definition removes the confounding created by mutation. However, mutation still sets the baseline for the information maintained. The definitions $I_{KL}$ and $I_{\Delta S}$ coincide if $p_i^\mu =  1/N \ \forall i$. In this case, the degree of localization or $F$ monotonically increases with the information maintained, with $I_{KL}=I_{\Delta S}=0$ when $F = 0$. For general $p_i^\mu$ that can be characterized by a localization factor $F_\mu$, $I_{KL}$ can be increased by setting $|F-F_\mu| > 0$. This means that the maintained information, through Eq. \ref{eq:info_kl}, can be increased potentially even by delocalization. For small mutation rates, this is not entirely surprising because selection may favor multiple types. But Eq. \ref{eq:info_kl} highlights that, for general mutation rates, a delocalized state may still be the result of strong variation in fitness.

\subsection{\label{sec:complexity}Viral complexity}

On the side of application, we note that Eq. \ref{eq:log_relation} also gives rise to a theory-driven measure for viral complexity: $\mathcal{H}(\textbf{p}^*)$ or its logarithm. Viral complexity has been defined as variation in a set of haplotypes quantified independently of the population size \cite{Gregori2016, Domingo2012}. Complexity indices essentially measure the quasispecies' diversity, which has implications for its adaptability \cite{Domingo2015, Pfeiffer2005, Vignuzzi2005} as well as disease progression and treatment \cite{Farci2000, Ojosnegros2008, Perales2011}. Some of the suggested complexity indices \cite{Gregori2016} include information-theoretic measures such as the Shannon entropy and diversity indices borrowed from ecology such as Hill numbers.

We propose that the logarithm of $\mathcal{H}(\textbf{p}^*)$ serve as an additional complexity index. Its unique advantage over other frequency-based measures is its direct connection with Eigen's model. Specifically, it carries an underlying dynamical, or biological, meaning given by Eq. \ref{eq:log_relation}. For example, suppose that for a given set of haplotype frequencies, we computed $\log_{10}{\mathcal{H}(\textbf{p}^*)} = 2$. If Eigen's model applies to the population, then Eq. \ref{eq:log_relation} implies that the standard deviation of effective fitness (i.e. the influx rate) of haplotypes is $10$ times larger than the mean mutation rate. But even if Eigen's model does not apply, $\log_b{\mathcal{H}(\textbf{p}^*)}$ is still useful as a metric. In any case, $\log{\mathcal{H}(\textbf{p}^*)}$ can be interpreted \textit{by analogy} with a population that evolves under Eigen's model. This is in the same vein that Hill numbers, which we introduce in the next paragraphs, can be interpreted by analogy with a population of types of equal abundance. 

Lastly, we note that $\mathcal{H}(\textbf{p}^*)$ can be written in terms of these ecological diversity metrics also proposed for viral complexity. The denominator of $\mathcal{H}(\textbf{p}^*)$ is the Gini-Simpson index. It is the probability of randomly picking two individuals of different types from the population. It makes sense that it is inversely proportional to $\bbV{s}$, which for large enough values, would concentrate the population on a single type. 

To interpret the numerator of $\mathcal{H}(\textbf{p}^*)$, we introduce Hill numbers. The Hill number of order $a$ is defined as ${}^a\!H := (\sum_{i=1}^Np_i^a)^{1/1-a}$ \cite{Hill1973}. It is meant to be a standardized diversity metric and can be interpreted as the effective number of equally abundant species (or types) that would give the same diversity ${}^a\!H$ \cite{Chao2014}. The order $a$ decides the weighting of the type frequencies in this metric. The Hill number of order $a = 0$ is simply the number of types $N$, placing equal weights on all type frequencies. When $a = -1$, which would yield the square root of the sum in the numerator, more weight is placed on rare (i.e. small frequency) types. Now, the Hill numbers are monotonically decreasing for increasing $a$. Negative orders would therefore produce diversity or effective numbers much greater than $N^2$. The numerator may then be interpreted as the ``excess diversity'' due to the presence of rare types. That this is inversely proportional to the mean mutation rate follows from the tendency of mutation to distribute the frequency mass and prevent such rare types. The metric $\mathcal{H}(\textbf{p}^*)$ can be fully expressed in terms of these Hill numbers:
\begin{equation}\label{eq:hill_relation}
    \mathcal{H}(\textbf{p}^*) = \dfrac{({}^{-1}\!H^*)^2-({}^0\!H^*)^2}{1-({}^{2}\!H^*)^{-1}},
\end{equation}
where the asterisk denotes that the Hill numbers are evaluated at equilibrium. 

We note that Hill numbers with negative orders are usually not computed in ecology. However, the Hill number of order $a=-1$ plays a clear and important role here. That said, the presence of inverse frequency terms would require the sampling effort to be comprehensive to detect rare haplotypes in actual populations. Higher-order Hill numbers, such as the Shannon entropy, which discount rare types more are still subject to sampling bias \cite{Gregori2014}. We admit that this does not make Eq. \ref{eq:hill_numbers} easily applicable to real data compared to traditional metrics. Nevertheless, Eq. \ref{eq:log_relation} provides an impetus to detect low-frequency haplotypes, which serves as a trade-off for a more biologically meaningful complexity index.

\section{\label{sec:conclusion}Summary and conclusion}

Next-generation sequencing coupled with algorithmic advances \cite{PosadaCespedes2017, Knyazev2020} have revealed the great heterogeneity of viral populations. Eigen's model, which introduced the quasispecies concept, has been very influential in understanding this mutation-driven diversity. But analytical results often rely on particular landscapes and mutation schemes or focus on extreme cases (e.g. error catastrophe or survival of the fittest). In this work, we derived a general relation, Eq. \ref{eq:log_relation}, which links the statistics of the model parameters to the population's equilibrium frequencies (Sec. \ref{sec:eq_relation}). We showed that the ratio of an effective fitness variance to the mean mutation rate squared, which we call the localization factor $F$, controls the degree of localization. For an initially quasilocalized population, we also obtained a critical value of $F$, Eq. \ref{eq:phase_retain} or Eq. \ref{eq:phase_cv}, that acts like a delocalization threshold (Sec. \ref{sec:critical_factor}). These relations hold as long as the number of types $N$ is large and the replication and mutation rates vary smoothly. Under these general assumptions, our results reinforce other observations \cite{Watkins2008, Peck2010, Hledk2022} that mutation impedes maintenance of information (Sec. \ref{sec:information}).

We have also proposed a viral complexity or diversity index (Sec. \ref{sec:complexity}). Compared to other such diversity metrics, Eq. \ref{eq:hill_numbers} or its logarithm is directly connected to Eigen's model and has a biological interpretation. However, because it depends on inverse frequency terms and on $N$ explicitly, it is more difficult to evaluate accurately compared to traditional measures. Still, rich high-throughput data paired with statistical models (e.g. \cite{Fuhrmann2024}) will prove very useful in this regard. Quantifying potential biases with simulated datasets \cite{Gregori2014} could be explored in a separate work. In any case, Eq. \ref{eq:hill_numbers} motivates accurate reconstruction of minority haplotypes. 

Our work adds to recent efforts applying speed limits or trade-off relations to population dynamics \cite{Zhang2020, Adachi2022, Hoshino2023, GarcaPintos2024}. Together with these studies, our results invite further applications of information theory or stochastic thermodynamics to biological systems. We concur with Ref. \cite{Hoshino2023}, for example, that the decomposition of the surprisal rate may yield insights in other models as it did here. For instance, Eigen's model can be extended to include recombination rates. It would be interesting to see how recombination modifies the localization factor. We leave these considerations for future work.

\begin{acknowledgments}
CJP thanks F. Tablizo for encouraging feedback and useful discussions. CJP was supported in part by grants to the Philippine Genome Center from the Department of Health - Epidemiology Bureau for the national SARS-CoV-2 biosurveillance initiative.
\end{acknowledgments}

\bibliography{bib}

\widetext
\clearpage

\appendix
\setcounter{page}{1}
\renewcommand{\thepage}{S\arabic{page}}

 \part*{ 
 \begin{center}
 \normalsize{
 Supplemental Material for ``Quasilocalization under coupled mutation-selection dynamics''
 } 
 \end{center}
 }

\begin{center}
C. J. Palpal-latoc\orcidlink{0000-0002-9171-6252}\\
\textit{
 Core Facility for Bioinformatics, Philippine Genome Center,\\ 
 University of the Philippines System, Quezon City 1101, Philippines and
}\\
\textit{
 National Institute of Physics, University of the Philippines Diliman, Quezon City 1101, Philippines
}
\end{center}

\begin{center}
Ian Vega\orcidlink{0000-0002-0428-048X}\\
\textit{
 National Institute of Physics, University of the Philippines Diliman, Quezon City 1101, Philippines
}
\end{center}

\section{Mean approximation of localization and delocalization speeds}

In the following derivations, the summation indices always run from one to $N$, subject to the indicated constraints. Sometimes, when there are three or more indices, only one summation symbol is used for brevity. But in any case, each index is always summed over. When taking the expectation of the sums, it is assumed that the rates are independent of the frequencies. This assumption breaks down in the localized regime. The relevant correction to the approximate localization speed is extracted in Sec. \ref{app:correction_terms}.

\subsection{Approximation of the localization speed}\label{app:loc_speed}

The square of the localization speed is given by
\begin{equation}
    (v_s)^2 = 
    \ser{i}p_i(\dot{\sigma}_i^s)^2 = \ser{i}s_i^2p_i - \lrpar{\ser{i}s_ip_i}^2.
\end{equation}
Then its expectation value is
\begin{align}
    \bbE{(\vsel)^2}
    &= \bbE{\ser{i}s_i^2p_i}-\bbE{\lrpar{\ser{i}s_ip_i}^2},\\
    &= \bbE{s^2}-\bbE{\ser{i}s_i^2p_i^2+\ser{i}\nser{j}{i}s_ip_is_jp_j},\\
    &= \bbE{s^2}-\bbE{s^2}\ser{i}p_i^2-\bbE{s_is_j}\ser{i}\nser{j}{i}p_ip_j,\label{app_eq:loc_means}\\
    &= \bbE{s^2}-\bbE{s^2}\ser{i}p_i^2-\bbE{s}^2\lrbra{1-\ser{i}p_i^2},\\
    &= \bbE{s^2}-\bbE{s^2}\ser{i}p_i^2 - \bbE{s}^2 + \bbE{s}^2\ser{i}p_i^2,\\
    &=  \lrpar{\bbE{s^2}-\bbE{s}^2}\lrpar{1-\ser{i}p_i^2}.
\end{align}
So,
\begin{equation}\label{app_eq:approx_loc_speed}
    \bbE{(\vsel)^2} = \bbV{s}\lrpar{1-\ser{i}p_i^2}.
\end{equation}

\subsection{Approximation of the delocalization speed}\label{app:deloc_speed}

The square of the delocalization speed is given by
\begin{align}
    (v_\mu)^2 &= 
    \ser{i}p_i(\dot{\sigma}_i^\mu)^2,\\
    &= \ser{i}\dfrac{1}{p_i}\lrpar{\ser{j}\mu_{ij}p_j}^2 - \lrpar{\ser{i}\ser{j}\mu_{ij}p_j}^2.
\end{align}
Extracting the diagonal mutation rate terms, we obtain
\begin{equation}
    (v_\mu)^2 = \ser{i} p_i\lrpar{\nser{j}{i}\mu_{ij}\dfrac{p_j}{p_i}}^2 - \lrpar{\ser{i}\nser{j}{i}\mu_{ij}p_j}^2 + 2\nser{j}{i}\mu_{ij}p_j\mu_{ii} + \ser{i}\mu_{ii}^2p_i -2\nser{j}{i}\mu_{ij}p_j\ser{k}\mu_{kk}p_k - \lrpar{\ser{i}\mu_{ii}p_i}^2 
\end{equation}

\subsubsection{Mean of off-diagonal mutation contributions}

Let us first take the mean of the off-diagonal terms. Note that
\begin{align}
    \ser{i}\dfrac{1}{p_i}\lrpar{\nser{j}{i}\mu_{ij}p_j}^2 &= \ser{i}\dfrac{1}{p_i}\lrpar{\nser{j}{i}\nser{k}{i}\mu_{ij}p_j\mu_{ik}p_k}, \\
    &= \ser{i}\dfrac{1}{p_i}\lrpar{\nser{j}{i}\mu_{ij}^2p_j^2} + \ntriser{j}{k}{i}\dfrac{1}{p_i}\mu_{ij}p_j\mu_{ik}p_k.
\end{align}
Then, 
\begin{align}
    \begin{split}
        &\bbE{\ser{i}\dfrac{1}{p_i}\lrpar{\nser{j}{i}\mu_{ij}p_j}^2}\\ 
        &=\bbE{\mu^2}\nser{j}{i}\dfrac{p_j^2}{p_i} + \bbE{\mu}^2\ntriser{j}{k}{i}\dfrac{p_jp_k}{p_i},
    \end{split}\\
    &=\bbE{\mu^2}\lrbra{\ser{i}\ser{j}\dfrac{p_j^2}{p_i}-1}+ \bbE{\mu}^2\lrpar{\ser{i}\nser{j}{i}\nser{k}{i}\dfrac{p_jp_k}{p_i} - \ser{i}\nser{j}{i}\dfrac{p_j^2}{p_i}},\\
    &= \bbE{\mu^2}\lrbra{\ser{i}\ser{j}\dfrac{p_j^2}{p_i}-1} + \bbE{\mu}^2\lrpar{\ser{i}\ser{j}\nser{k}{i}\dfrac{p_jp_k}{p_i}-\ser{i}\nser{k}{i}p_k} - \bbE{\mu}^2\lrpar{\ser{i}\ser{j}\dfrac{p_j^2}{p_i} - 1},\\
    &= \bbV{\mu}\lrpar{\ser{i}\ser{j}\dfrac{p_j^2}{p_i}-1} + \bbE{\mu}^2\lrpar{\ser{i}\ser{j}\nser{k}{i}\dfrac{p_jp_k}{p_i}
    -\ser{i}\nser{k}{i}p_k},\\
    &= \bbV{\mu}\lrpar{\ser{i}\ser{j}\dfrac{p_j^2}{p_i}-1}+ \bbE{\mu}^2\lrpar{\ser{i}\nser{k}{i}\dfrac{p_k}{p_i}-\ser{i}\nser{k}{i}p_k},\\
    &=\bbV{\mu}\lrpar{\ser{i}\ser{j}\dfrac{p_j^2}{p_i}-1}+ \bbE{\mu}^2\lrpar{\ser{i}\dfrac{1-p_i}{p_i}-\ser{i}\lrpar{1-p_i}},\\
    &= \bbV{\mu}\lrpar{\ser{i}\ser{j}\dfrac{p_j^2}{p_i}-1}+ \bbE{\mu}^2\lrpar{\ser{i}\dfrac{1}{p_i}-N-N+1},\\
    &= \bbV{\mu}\lrpar{\ser{i}\ser{j}\dfrac{p_j^2}{p_i}-1}+ \bbE{\mu}^2\lrpar{\ser{i}\dfrac{1}{p_i}-2N+1},\\
    &= \bbV{\mu}\nser{j}{i}\dfrac{p_j^2}{p_i}+ \bbE{\mu}^2\lrpar{\ser{i}\dfrac{1}{p_i}-2N+1}.
\end{align}
Next,
\begin{equation}
    \bbE{\lrpar{\ser{i}\nser{j}{i}\mu_{ij}p_j}^2},\\
    = \bbE{\ser{i}\nser{j}{i}\ser{k}\nser{l}{k}\mu_{ij}p_j\mu_{kl}p_l}.
\end{equation}
Now, $\mu_{ij}=\mu_{kl}$ only when $(i, j)=(k,l)$. We can write the sum above as
\begin{align}
    \bbE{\Ave{\mu}^2} &= \bbE{\ser{i}\nser{j}{i}\mu_{ij}^2p_j^2} + \bbE{\sum_{\substack{(i,j)\ne (k,l)\\ j\ne i\\ l\ne k}}\mu_{ij}p_j\mu_{kl}p_l}\\
    &= \bbE{\mu^2}\ser{i}\nser{j}{i}p_j^2 + \bbE{\mu}^2\sum_{\substack{(i,j)\ne (k,l)\\ j\ne i\\ l\ne k}}p_jp_l,\\
    &= \bbE{\mu^2}\ser{i}\nser{j}{i}p_j^2 + \bbE{\mu}^2\lrbra{\lrpar{\ser{i}\nser{j}{i}p_j}^2-\ser{i}\nser{j}{i}p_j^2},\\
    &= \bbV{\mu}\ser{i}\nser{j}{i}p_j^2 + \bbE{\mu}^2\lrpar{\ser{i}\nser{j}{i}p_j}^2,\\
    &=  \bbV{\mu}\ser{i}\nser{j}{i}p_j^2 + \bbE{\mu}^2\lrpar{\ser{i}(1-p_i)}^2,\\
    &= \bbV{\mu}\ser{i}\nser{j}{i}p_j^2 + \bbE{\mu}^2\lrpar{N-1}^2.
\end{align}
Then, the total off-diagonal contribution is
\begin{align}
    \bbE{(\vmut)^2}\bigg|_\text{off-diagonal} &= \bbV{\mu}\nser{j}{i}\dfrac{p_j^2}{p_i}+ \bbE{\mu}^2\lrpar{\ser{i}\dfrac{1}{p_i}-2N+1} - \lrbra{\bbV{\mu}\ser{i}\nser{j}{i}p_j^2 + \bbE{\mu}^2\lrpar{N-1}^2},\\
    &= \bbE{\mu}^2\lrbra{\ser{i}\dfrac{1}{p_i}-2N+1 - \lrpar{N-1}^2} + \bbV{\mu}\ser{i}\nser{j}{i}\lrpar{\dfrac{p_j^2}{p_i}-p_j^2}.
\end{align}
Simplifying,
\begin{equation}
    \bbE{(\vmut)^2}\bigg|_\text{off-diagonal} = \bbE{\mu}^2\lrpar{\ser{i}\dfrac{1}{p_i}-N^2} + \bbV{\mu}\ser{i}\nser{j}{i}\lrpar{\dfrac{1}{p_i}-1}p_j^2.
\end{equation}

\subsubsection{Mean of diagonal mutation contributions}

The diagonal contributions to the delocalization speed are
\begin{align}
   (\vmut)^2\bigg|_\text{diagonal} &= 2\ser{i}\nser{j}{i}\mu_{ij}p_j\mu_{ii} + \ser{i}\mu_{ii}^2p_i -2\lrpar{\ser{i}\nser{j}{i}\mu_{ij}p_j}\ser{k}\mu_{kk}p_k - \lrpar{\ser{i}\mu_{ii}p_i}^2,\\
   \begin{split}
       &= -2\ser{i}\nser{j}{i}\nser{k}{i}\mu_{ij}\mu_{ik}p_j + \ser{i}\nser{j}{i}\nser{k}{i}\mu_{ij}\mu_{ik}p_i,\\
       &\quad+2\ser{i}\nser{j}{i}\ser{k}\nser{l}{k}\mu_{ij}\mu_{kl}p_jp_k - \ser{i}\nser{j}{i}\ser{k}\nser{l}{k}\mu_{ij}\mu_{kl}p_ip_k.
   \end{split}
\end{align}
Let us first tackle the triple sums, which share the same mean, except for the factor $-2$. Expanding one of them, we have
\begin{equation}
    \ser{i}\mu_{ii}^2p_i = \ser{i}\nser{j}{i}\mu_{ij}^2p_j + \ntriser{i}{j}{k}\mu_{ij}\mu_{ik}p_i.
\end{equation}
Then,
\begin{align}
    \bbE{\ser{i}\mu_{ii}^2p_i} &= \bbE{\mu^2}\ser{i}\nser{j}{i}p_i + \bbE{\mu}^2\ntriser{i}{j}{k}p_i,\\
    &= \bbE{\mu^2}(N-1) + \bbE{\mu}^2(N-2)(N-1),\\
    &= \bbE{\mu^2}(N-1) + \bbE{\mu}^2\lrbra{(N-1)^2-(N-1)},\\
    &= \bbV{\mu}(N-1) + \bbE{\mu}^2(N-1)^2.
\end{align}
Thus, the mean of the two triple sums is
\begin{equation}
    \bbE{2\ser{i}\nser{j}{i}\mu_{ii}\mu_{ij}p_j + \ser{i}\mu_{ii}^2p_i} = -\bbV{\mu}(N-1) - \bbE{\mu}^2(N-1)^2.
\end{equation}

On the other hand, the pair of quadruple sums do not share the same mean. Expanding one of them, we have
\begin{equation}
    \lrpar{\ser{i}\mu_{ii}p_i}^2 = \ser{i}\nser{j}{i}\mu_{ij}^2p_i^2 + \sum_{\substack{(i,j)\ne(k,l)\\j\ne i\\k\ne l)}}\mu_{ij}\mu_{kl}p_ip_k.
\end{equation}
Then,
\begin{align}
    \bbE{\lrpar{\ser{i}\mu_{ii}p_i}^2} &= \bbE{\mu^2}\ser{i}\nser{j}{i}p_i^2 + \bbE{\mu}^2\sum_{\substack{(i,j)\ne(k,l)\\j\ne i\\k\ne l)}}p_ip_k,\\
    &=\bbE{\mu^2}(N-1)\ser{i}p_i^2 + \bbE{\mu}^2\sum_{\substack{(i,j)\ne(k,l)\\j\ne i\\k\ne l)}}p_ip_k,\\
    &=\bbE{\mu^2}(N-1)\ser{i}p_i^2 + \bbE{\mu}^2\lrpar{\ser{i}\nser{j}{i}\ser{k}\nser{l}{k}p_ip_k - \ser{i}\nser{j}{i}p_i^2},\\
    &= \bbE{\mu^2}(N-1)\ser{i}p_i^2 + \bbE{\mu}^2\lrbra{(N-1)^2 -(N-1)\ser{i}p_i^2},\\
    &= \bbV{\mu}(N-1)\ser{i}p_i^2 + \bbE{\mu}^2(N-1)^2.
\end{align}
Next,
\begin{equation}
    \ser{i}\nser{j}{i}\ser{k}\mu_{ij}p_j\mu_{kk}p_k = \ser{i}\nser{j}{i}\mu_{ij}^2p_jp_i + \sum_{\substack{(i,j)\ne(k,l)\\j\ne i\\k\ne l)}}\mu_{ij}\mu_{kl}p_jp_k.
\end{equation}
Then,
\begin{align}
    \bbE{\ser{i}\nser{j}{i}\ser{k}\mu_{ij}p_j\mu_{kk}p_k} &= \bbE{\mu^2}\ser{i}\nser{j}{i}p_jp_i + \bbE{\mu}^2\sum_{\substack{(i,j)\ne(k,l)\\j\ne i\\k\ne l)}}p_jp_k,\\
    &= \bbE{\mu^2}\lrpar{1-\ser{i}p_i^2} + \bbE{\mu}^2\sum_{\substack{(i,j)\ne(k,l)\\j\ne i\\k\ne l)}}p_jp_k,\\
    &= \bbE{\mu^2}\lrpar{1-\ser{i}p_i^2} + \bbE{\mu}^2\lrbra{\ser{i}\nser{j}{i}\ser{k}\nser{l}{k}p_ip_k - \lrpar{1-\ser{i}p_i^2}},\\
    &= \bbV{\mu}\lrpar{1-\ser{i}p_i^2} + \bbE{\mu}^2(N-1)^2.
\end{align}
So the mean of the two quadruple sums is
\begin{equation}
    \bbE{ -2\lrpar{\ser{i}\nser{j}{i}\mu_{ij}p_j}\ser{k}\mu_{kk}p_k - \lrpar{\ser{i}\mu_{ii}p_i}^2} = \bbV{\mu}\lrbra{-\ser{i}p_i^2(N+1)-2} + \bbE{\mu}^2(N-1)^2.
\end{equation}
The total diagonal contribution is
\begin{equation}
    \bbE{(\vmut)^2}\bigg|_\text{diagonal} = \bbV{\mu}\lrbra{-\ser{i}p_i^2(N+1)-(N-1)+2}.
\end{equation}

\subsubsection{Total mean of the delocalization speed}

Combining the diagonal and off-diagonal contributions, the approximate delocalized speed (squared) is
\begin{equation}
    \bbE{(v_\mu)^2} = \bbE{\mu}^2\lrpar{\ser{i}\dfrac{1}{p_i}-N^2} + \bbV{\mu}\lrbra{\ser{i}p_i^2\lrpar{\ser{j}\dfrac{1}{p_j} - 2N} -N+2}.
\end{equation}

\section{Equality of speeds at equilibrium}\label{app:equal_speeds}

We show explicitly that the localization and delocalization speeds equal each other at equilibrium. The eigenvalue equation states
\begin{equation}
    \nser{j}{i}\mu_{ij}h_{j\text{max}} = \lambda_\text{max}h_{i\text{max}} - W_{ii}h_{i\text{max}}.
\end{equation}
With $h_\text{tot} := \ser{i}h_{i\text{max}}$,
\begin{equation}
    (v_\mu)^2 = \ser{i}\dfrac{1}{h_{i\text{max}h_\text{tot}}}\lrpar{\lambda_\text{max}h_{i\text{max}}-W_{ii}h_{i\text{max}} + \mu_{ii}h_{i\text{max}}}^2 - \dfrac{1}{h_\text{tot}^2}\lrbra{\ser{k}\lrpar{\lambda_\text{max}h_{k\text{max}}-W_{kk}h_{k\text{max}}} + \ser{k}\mu_{kk}h_{k\text{max}}}^2
\end{equation}
But $W_{ii} = s_i + \mu_{ii}$. So, 
\begin{align}
    (v_\mu)^2 &= \ser{i}\dfrac{1}{h_{i\text{max}}h_\text{tot}}\lrpar{\lambda_\text{max} -s_ih_{i\text{max}}}^2 - \dfrac{1}{h_\text{tot}^2}\lrpar{\ser{k}\lambda_\text{max}h_{k\text{max}} - \ser{k}s_kh_{k\text{max}}}^2,\\
    &= \ser{i}\dfrac{h_{i\text{max}}}{h_\text{tot}}\lrpar{\lambda_\text{max}-s_i}^2 - \dfrac{1}{h_\text{tot}^2}\lrbra{\ser{k}\lrpar{\lambda_\text{max}-s_k}h_{k\text{max}}}^2,\\
    &= \ser{i}\dfrac{h_{i\text{max}}}{h_\text{tot}}\lrpar{\lambda_\text{max}^2 - 2\lambda_\text{max}s_i + s_i^2} - \dfrac{1}{h_\text{tot}^2}\lrbra{\lrpar{\ser{k}\lambda_\text{max}h_{k\text{max}}}^2 -2\ser{k}\lambda_\text{max}h_{k\text{max}}\ser{k}s_kh_{k\text{max}} + \lrpar{\ser{k}s_kh_{k\text{max}}}^2},\\
    &= \ser{i}p_i^*\lrpar{\lambda_\text{max}^2 - 2\lambda_\text{max}s_i + s_i^2} - \lrbra{\lrpar{\ser{k}\lambda_\text{max}p_k^*}^2 -2\ser{k}\lambda_\text{max}p_k^*\ser{k}s_kp_k^* + \lrpar{\ser{k}s_kp_k^*}^2},\\
    &= \lambda_\text{max}^2 - 2\lambda_\text{max}\Ave{s} + \Ave{s^2} - \lambda_\text{max}^2 + 2\lambda_\text{max}\Ave{s}-\Ave{s}^2,\\
    &= \Ave{s^2} - \Ave{s}^2.
\end{align}
Thus, $v_\mu = v_s$ at equilibrium.

\section{Numerical implementation and additional results}\label{app:numerical_implementation}

The full code to reproduce the figures or calculations is available for download at \cite{palpallatoc}. Here, we briefly mention how the rates are drawn and how a single realization of a quasispecies population is simulated for a given localization factor $F$. To obtain the quasispecies $\textbf{p}^*$, we need only solve for the maximum eigenvalue and the associated right eigenvector of the growth matrix $\mathsf{W}$. To build $\mathsf{W}$, we need to draw $N$ influx rates $s_i$ and $N^2 - N$ mutation rates $\mu_{ij} \ (j\ne i)$. 

\begin{figure}
    \centering
    \includegraphics[width=0.7\linewidth]{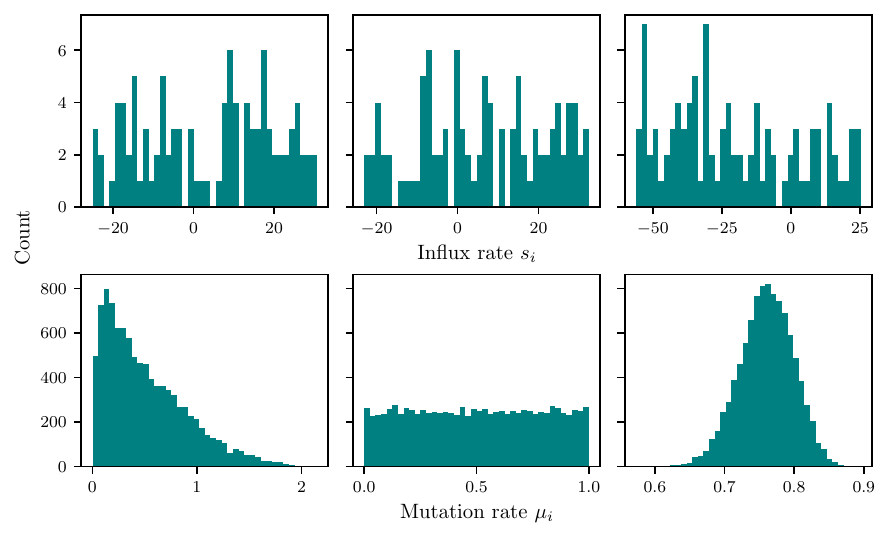}
    \caption{Histograms (bin size is $40$) of influx (top) and mutation rates (bottom). Each column shows the rate distributions  from one realization with $F = 10^3$ and $N = 100$. The left column shows mutation rates drawn from the Dirichlet distribution, the middle column from the uniform distribution, and the right column from the beta distribution.}
    \label{fig:supp_distributions}
\end{figure}

In this work, the influx rates are always drawn from a uniform distribution. These rates may be positive or negative (if the mutation rates are much larger or because the degradation rates are large enough). Meanwhile, the mutation rates, which are never negative, may be drawn from the (1) Dirichlet distribution, (2) the uniform distribution, and the (3) beta distribution. To draw mutation rates from the Dirichlet distribution, we draw $N$ Dirichlet vectors (of size $N-1$) with uniform-random parameters. Next, we draw $N$ random numbers from a uniform distribution that we use to variably scale each of the vectors. Sampling is done this way to mimic the constraint $\sum_{i=1}^NQ_{ji}A_i = A_i$. (Note that this does not impose a constraint on the influx rates.) To draw from the beta distribution, the distribution parameters are drawn uniformly first. Drawing from the uniform distribution is straightforward. In Fig. \ref{fig:supp_distributions}, we show examples of sampled rates across different distributions.

\begin{figure}
    \centering
    \includegraphics[width=0.7\linewidth]{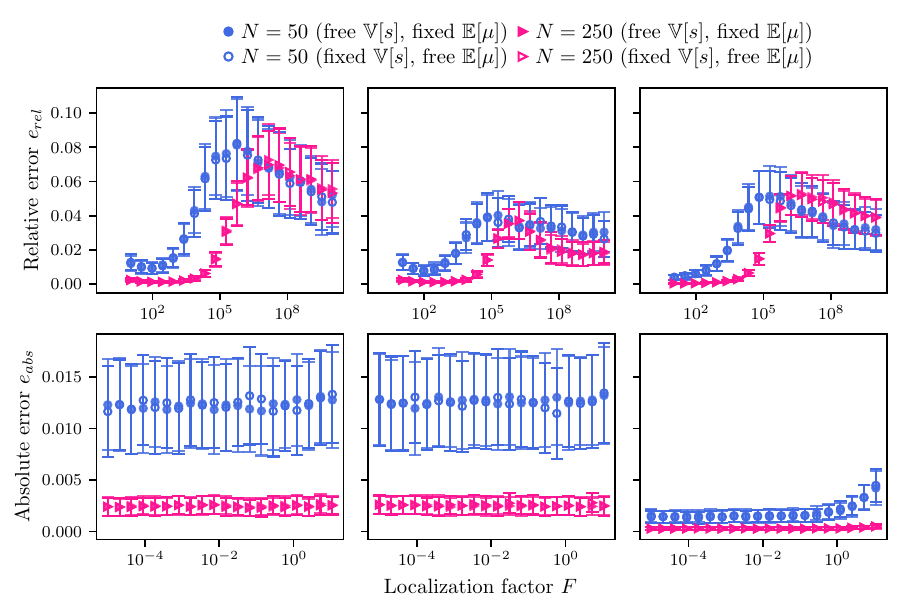}
    \caption{Accuracy of Eq. \ref{eq:log_relation} when influx rates may be negative or positive. Each column shows relative and absolute errors when mutation rates are drawn from the Dirichlet distribution (left column), the uniform distribution (middle), and the beta distribution (right). The colors denote the number of types. Open markers indicate the errors when the influx rate variance is held fixed (here, $\bbV{s} =  100$) but the mean mutation rate is varied to get the desired $F$. Meanwhile, filled markers show the opposite, when the influx rate variance is allowed to vary to achieve a given $F$.}
    \label{fig:supp_logerrors_nonzero}
\end{figure}

\begin{figure}
    \centering
    \includegraphics[width=0.7\linewidth]{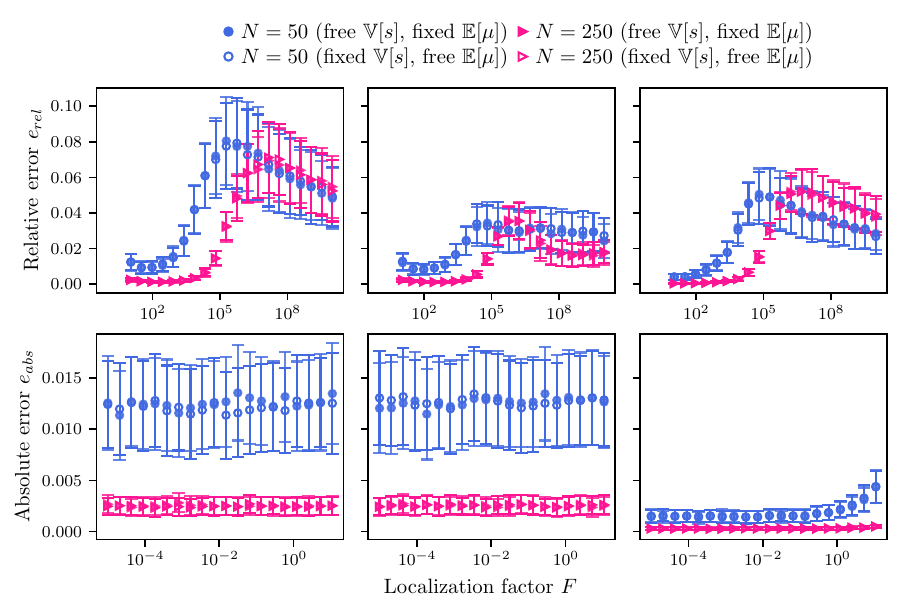}
    \caption{Accuracy of Eq. \ref{eq:log_relation} when all influx rates are positive. Colors, markers, and columns are as detailed in Fig. \ref{fig:supp_logerrors_nonzero}.}
    \label{fig:supp_logerrors_positive}
\end{figure}

To generate a quasispecies for a given $F$, we need to adjust the distribution parameters of the rates to get the desired $\bbV{s}/\bbE{\mu}^2 =: F$. When simulating across different factors, we can either vary $\bbV{s}$ or set it to a fixed constant and vary $\bbE{\mu}$ instead. When $\bbV{s}$ is allowed to vary, we draw mutation rates with their default parameters and then adjust $\bbV{s}$ accordingly. In this case, the mean mutation rate $\bbE{\mu}$ is fixed except when drawing from the beta distribution whose default parameters are random and therefore does not have the same skewness at each instance. When $\bbV{s}$ is fixed instead, $\bbE{\mu}$ is adjusted. In Figs. \ref{fig:supp_logerrors_nonzero} and \ref{fig:supp_logerrors_positive}, we show the relative and absolute errors of Eq. \ref{eq:log_relation} across all three mutation rate distributions, whether the influx rates are all positive or may be negative, and whether $\bbV{s}$ is fixed or adjusted directly. Overall, we find similar results with relative errors dropping below $6\%$ outside the lcoalized regime. 

\FloatBarrier

\section{Covariance correction terms}\label{app:correction_terms}

\subsection{Analytical correction to the localization speed approximation}

We extract the correction terms to each of the three approximations in Eq. \ref{app_eq:loc_means}. First,
\begin{equation}
    \ser{i}s_i^2p_i = \bbE{s^2}\ser{i}p_i + \ser{i}p_i\lrpar{s_i^2-\bbE{s^2}}.
\end{equation}
Then, the error in the first mean is
\begin{equation}
    \ser{i}p_i\lrpar{s_i^2-\bbE{s^2}}  = \ser{i}s_i^2p_i - \bbE{s^2}\ser{i}p_i.
\end{equation}
This is simply the covariance between $s$ and $p$:
\begin{equation}
    \dfrac{1}{N}\ser{i}p_i\lrpar{s_i^2-\bbE{s^2}} = \bbE{s^2p} - \bbE{s^2}\bbE{p} = \bbC{s^2}{p}.
\end{equation}
Next, 
\begin{equation}
    \ser{i}s_i^2p_i^2 = \bbE{s^2}\ser{i}p_i^2 + \ser{i}p_i^2\lrpar{s_i^2-\bbE{s^2}}.
\end{equation}
The error in the second mean is
\begin{equation}
    \ser{i}p_i^2\lrpar{s_i^2-\bbE{s^2}} = \ser{i}s_i^2p_i^2 - \bbE{s^2}\ser{i}p_i^2.
\end{equation}
This is also a covariance but between $s^2$ and $p^2$:
\begin{equation}
    \dfrac{1}{N}\ser{i}p_i^2\lrpar{s_i^2-\bbE{s^2}} = \bbE{s^2p^2} - \bbE{s^2}\bbE{p^2} = \bbC{s^2}{p^2}.
\end{equation}
For the last mean, we define $\Delta s_i := s_i - \bbE{s}$. Then,
\begin{align}
    \ser{i}\nser{j}{i}s_ip_is_jp_j &= \ser{i}\nser{j}{i}p_ip_j\lrpar{\Delta s_i + \bbE{s}}\lrpar{\Delta s_j + \bbE{s}},\\
    &= \ser{i}\nser{j}{i}p_ip_j\lrbra{\Delta s_i\Delta s_j + \bbE{s}(\Delta s_i + \Delta s_j) + \bbE{s}^2}.
\end{align}
So the error in the third mean is
\begin{equation}
    \ser{i}\nser{j}{i}s_ip_is_jp_j - \bbE{s}^2\ser{i}\nser{j}{i}p_ip_j = \ser{i}\nser{j}{i}p_ip_j\lrbra{\Delta s_i \Delta s_j + \bbE{s}\lrpar{\Delta s_i + \Delta s_j}}.
\end{equation}
We can rewrite the two parts of this error separately. First,
\begin{align}
    \ser{i}\nser{j}{i}p_ip_j\Delta s_i \Delta s_j &= \ser{i}\ser{j}p_i\Delta s_ip_j\Delta s_j - \ser{i}p_i^2\Delta s_i^2,\\
    &= N^2 \bbE{p\Delta s}^2 - N\bbE{p\Delta s},\\
    &= N^2\lrpar{\bbE{sp}-\bbE{s}\bbE{p}}^2 - N\lrpar{\bbE{sp}-\bbE{s}\bbE{p}}^2,\\
    &= N^2\bbC{s}{p}^2 - N\bbC{s}{p}^2.
\end{align}
Second,
\begin{align}
    \bbE{s}\ser{i}\nser{j}{i}p_ip_j\lrpar{\Delta s_i + \Delta s_j} &= 2\bbE{s}\ser{i}\nser{j}{i}p_i\Delta s_ip_j,\\
    &= 2\bbE{s}\ser{i}p_i\Delta s_i\lrpar{1-p_i},\\
    &= 2\bbE{s}N\lrpar{\bbE{p\Delta s} -  \bbE{p^2\Delta s}},\\
    &= 2\bbE{s}N\lrpar{\bbC{s}{p} - \bbC{s}{p^2}}.
\end{align}
The total correction to the approximation of the localization speed is
\begin{align}
    (v_s)^2 - \bbV{s}(1-\ser{i}p_i^2) &= N\bbC{s^2}{p} - N\bbC{s^2}{p^2} -N^2\bbC{s}{p}^2 + N\bbC{s}{p}^2-2\bbE{s}N\lrpar{\bbC{s}{p} - \bbC{s}{p^2}},\\
    &= -N^2\bbC{s}{p}^2-2\bbE{s}N\lrpar{\bbC{s}{p} - \bbC{s}{p^2}} + N\lrpar{\bbC{s^2}{p} -\bbC{s^2}{p^2}} + N\bbC{s}{p}^2.\label{app_eq:correction}
\end{align}
The middle two terms which involve covariance differences should be roughly on the same order. The covariances involving $s^2$ is larger than those only involving $s$, but this is compensated by the $\bbE{s}$ factor. Meanwhile, the $N^2$ term swamps the last covariance. Hence, the correction leans to the negative. 

\subsection{Numerical evaluation of the covariance corrections}\label{app:num_corrections}

\begin{figure}
    \centering
    \includegraphics[width=0.7\linewidth]{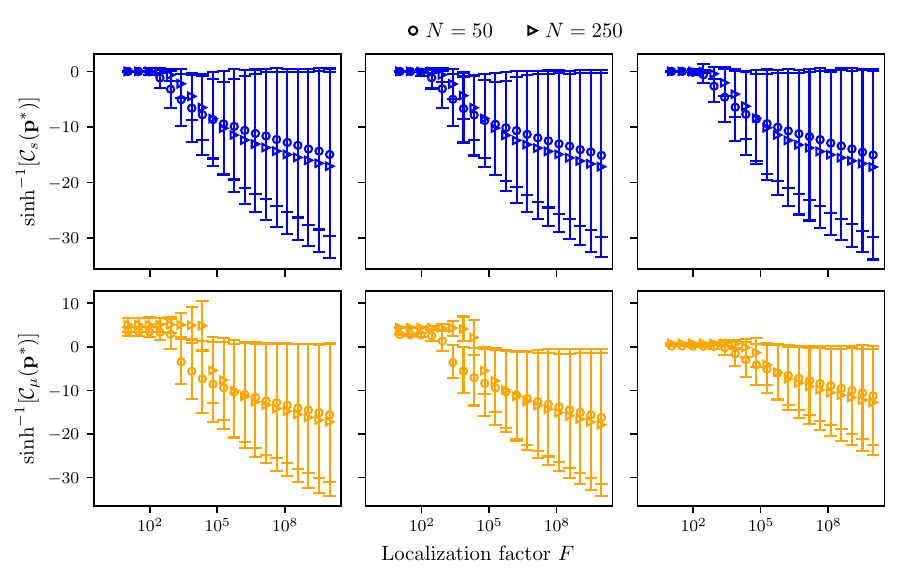}
    \caption{Growth of the covariance corrections. In the localized regime, the corrections are negative.  The top row shows the median corrections to the localization speed and the bottom row shows the corrections to the delocalization speed. From left to right, each column shows the corrections with mutation rates drawn from the Dirichlet distribution, the uniform distribution, and the beta distribution, respectively. Each point represents the median correction over $10^3$ realizations and the error bars cover the interquartile range.}
    \label{fig:supp_covcorrections}
\end{figure}

\begin{figure}
    \centering
    \includegraphics[width=0.7\linewidth]{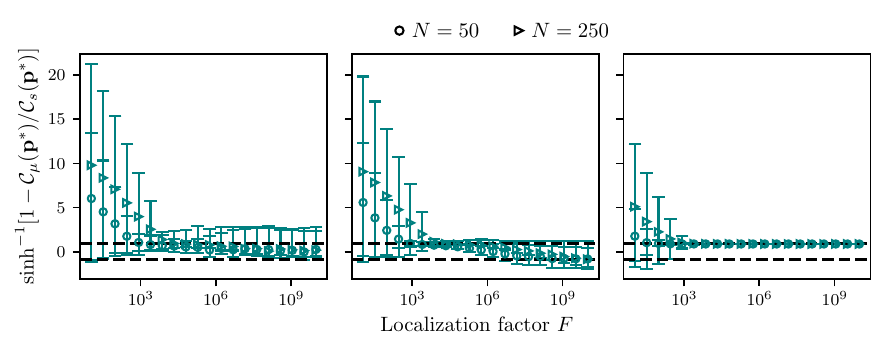}
    \caption{Comparison of the relative sizes of the corrections to the speed approximations. For small $F$, $\mathcal{C}_s(\mathbf{p}^*)$ remains largely dominant over $\mathcal{C}_\mu(\mathbf{p}^*)$ until the population enters the localized regime. The horizontal lines mark the range in which either correction is within $100\%$ of each other's values. Each plot shows the median relative error between $\mathcal{C}_s(\mathbf{p}^*)$ and $\mathcal{C}_\mu(\mathbf{p}^*)$ with mutation rates drawn from the Dirichlet distribution (left), the uniform distribution (middle), and the beta distribution (right), respectively. Each point represents the median error over $10^3$ realizations and the error bars cover the interquartile range.}
    \label{fig:supp_covcorr_relsize}
\end{figure}

In Fig. \ref{fig:supp_covcorrections}, we numerically evaluate the corrections to the speed approximations, $\mathcal{C}_s(\mathbf{p}^*) := v_s - \bbE{v_s}$ and $\mathcal{C}_\mu(\mathbf{p}^*) := v_\mu - \bbE{v_\mu}$, at equilibrium. We find that by and large the approximations overestimate the speeds and grow in magnitude as $F$ increases. Therefore, both $\mathcal{C}_s(\mathbf{p}^*)$ and $\mathcal{C}_\mu(\mathbf{p}^*)$ largely have negative signs in the localized regime (i.e. large $F$). In Fig. \ref{fig:supp_covcorr_relsize}, the correction $\mathcal{C}_\mu(\mathbf{p}^*)$ catches up to $\mathcal{C}_s(\mathbf{p}^*)$ within a $100\%$ relative error margin. Since both corrections have the same negative sign, the overall error in the equilibrium relation Eq. \ref{eq:log_relation} is markedly diminished as they cancel each other to a substantial degree. This explains the drop in error in Fig. \ref{fig:errors} in the localized regime. 

\FloatBarrier

\section{Critical \texorpdfstring{$s_\text{max}/\mu_\text{max}$}{smax/μmax} for uniformly distributed rates and initial frequencies}\label{app:uniform_dist_freq}

For an initially quasilocalized population and to leading order of $p$, we can rewrite Eq. \ref{eq:phase_retain} as 
\begin{equation}\label{app_eq:phase_retain}
    \dfrac{\mathbb{V}[s]}{\mathbb{E}[\mu]^2} \approx \sum_{i=1}^N\dfrac{1}{p_i}-N^2.
\end{equation}
A random variable $X$ with uniform distribution $U(a, b)$ has mean $\mathbb{E}[X] = (a+b)/2$ and variance $\mathbb{V}[X] = (b-a)^2/12$. Following the assumption in Ref. \cite{Hoshino2023} that $s \sim U(-s_\text{max}, s_\text{max})$ and $\mu \sim U(0, \mu_\text{max})$, we then have
\begin{equation}
    \dfrac{\mathbb{V}[s]}{\mathbb{E}[\mu]^2} = \dfrac{(4/12)}{(1/4)}\dfrac{s_\text{max}^2}{\mu_\text{max}^2} = \dfrac{4}{3}\dfrac{s_\text{max}^2}{\mu_\text{max}^2}.
\end{equation}
Plugging this back into Eq. \ref{app_eq:phase_retain}, we have
\begin{equation}
    \dfrac{s_\text{max}^2}{\mu_\text{max}^2} \approx \dfrac{3}{4}\left(\sum_{i=1}^N\dfrac{1}{p_i}-N^2\right).\label{eq:ratio_a}
\end{equation}
It remains to be shown that the right-hand side of the above equation is on the order of $N^2$. In Ref. \cite{Hoshino2023}, the distributions of the type population sizes have not been specified except that they are uniform. Therefore, we consider the general case that $n \sim U(a,b)$, where $a$ and $b$ are positive numbers. Making this general assumption, we then take the expectation value of the sum $\sum_i 1/p$. We can rewrite this sum explicitly in terms of the absolute frequencies as
\begin{equation}
    \sum_{i=1}^N\dfrac{1}{p_i} = \sum_{i=1}^N\sum_{j=1}^N \dfrac{n_j}{n_i} = N + \sum_{i=1}^N\sum_{j\ne i}^N\dfrac{n_j}{n_i}.
\end{equation}
Then,
\begin{align}
    \bbE{\sum_{i=1}^N\dfrac{1}{p_i}} &= N + N(N-1)\bbE{\dfrac{n_j}{n_i}},\\
    &= N + N(N-1)\bbE{n}\bbE{\dfrac{1}{n}},
\end{align}
where we made use of the fact that since $j\ne i$, then $n_j \ne n_i$ such that $n_j$ and $1/n_i$ are independent. For $n \sim U(a,b)$,
\begin{equation}
    \bbE{\dfrac{1}{n}} = \int_a^b\dfrac{1}{n}\dfrac{1}{b-a}dn = \dfrac{\ln{b/a}}{b-a}.
\end{equation}
So,
\begin{equation}
    \bbE{\sum_{i=1}^N\dfrac{1}{p_i}} = N + (N^2-N)\dfrac{1}{2}\dfrac{b+a}{b-a}\ln{b/a}.
\end{equation}
Because some types are in general fitter and others less so, the differences in the population sizes of the types can be orders of magnitude. Hence, it is better to consider the orders of $a$ and $b$ explicitly. If we let $a = 10^x$ and $b = 10^y$, where $y$ and $x$ are positive reals and $y > x$, then
\begin{equation}
    \dfrac{1}{2}\dfrac{b+a}{b-a}\ln{b/a} = \dfrac{1 + 10^{x-y}}{1 - 10^{x-y}}\dfrac{\ln 10}{2}(y-x).
\end{equation}
Eq. \ref{eq:ratio_a} can therefore be written as
\begin{equation}
    \dfrac{s_\text{max}^2}{\mu_\text{max}^2} = \dfrac{3}{4}(N^2-N)\lrbra{\dfrac{\ln 10}{2}\dfrac{1 + 10^{x-y}}{1 - 10^{x-y}}(y-x) - 1}. \label{eq:proof_b}
\end{equation}
For $y-x=2$, Eq. \ref{eq:proof_b} reduces to
\begin{align}
    \dfrac{s_\text{max}^2}{\mu_\text{max}^2} &\approx 0.75\times1.35(N^2 - N) \approx N^2,\\
    \implies \dfrac{s_\text{max}}{\mu_\text{max}} &\approx N.
\end{align}

\end{document}